\documentclass{article}
\usepackage{a4wide}
\usepackage{TRtitlepage}

\makeatletter
\@ifundefined{lhs2tex.lhs2tex.sty.read}%
  {\@namedef{lhs2tex.lhs2tex.sty.read}{}%
   \newcommand\SkipToFmtEnd{}%
   \newcommand\EndFmtInput{}%
   \long\def\SkipToFmtEnd#1\EndFmtInput{}%
  }\SkipToFmtEnd

\newcommand\ReadOnlyOnce[1]{\@ifundefined{#1}{\@namedef{#1}{}}\SkipToFmtEnd}
\usepackage{amstext}
\usepackage{amssymb}
\usepackage{stmaryrd}
\DeclareFontFamily{OT1}{cmtex}{}
\DeclareFontShape{OT1}{cmtex}{m}{n}
  {<5><6><7><8>cmtex8
   <9>cmtex9
   <10><10.95><12><14.4><17.28><20.74><24.88>cmtex10}{}
\DeclareFontShape{OT1}{cmtex}{m}{it}
  {<-> ssub * cmtt/m/it}{}

\DeclareFontShape{OT1}{cmtt}{bx}{n}
  {<5><6><7><8>cmtt8
   <9>cmbtt9
   <10><10.95><12><14.4><17.28><20.74><24.88>cmbtt10}{}
\DeclareFontShape{OT1}{cmtex}{bx}{n}
  {<-> ssub * cmtt/bx/n}{}

\newcommand{\Conid}[1]{\mathit{#1}}
\newcommand{\Varid}[1]{\mathit{#1}}
\newcommand{\anonymous}{\kern0.06em \vbox{\hrule\@width.5em}}

\newcommand{\bind}{\mathbin{>\!\!\!>\mkern-6.7mu=}}

\usepackage{polytable}

\@ifundefined{mathindent}%
  {\newdimen\mathindent\mathindent\leftmargini}%
  {}%

\def\resethooks{%
  \global\let\SaveRestoreHook\empty
  \global\let\ColumnHook\empty}
\newcommand*{\savecolumns}[1][default]%
  {\g@addto@macro\SaveRestoreHook{\savecolumns[#1]}}
\newcommand*{\restorecolumns}[1][default]%
  {\g@addto@macro\SaveRestoreHook{\restorecolumns[#1]}}
\newcommand*{\aligncolumn}[2]%
  {\g@addto@macro\ColumnHook{\column{#1}{#2}}}

\resethooks

\newcommand{\onelinecommentchars}{\quad-{}- }
\newcommand{\commentbeginchars}{\enskip\{-}
\newcommand{\commentendchars}{-\}\enskip}

\newcommand{\visiblecomments}{%
  \let\onelinecomment=\onelinecommentchars
  \let\commentbegin=\commentbeginchars
  \let\commentend=\commentendchars}

\newcommand{\invisiblecomments}{%
  \let\onelinecomment=\empty
  \let\commentbegin=\empty
  \let\commentend=\empty}

\visiblecomments

\newlength{\blanklineskip}
\newcommand{\hsindent}[1]{\quad}
\let\hspre\empty
\let\hspost\empty

\EndFmtInput
\makeatother
\ReadOnlyOnce{polycode.fmt}%
\makeatletter

\newcommand{\hsnewpar}[1]%
  {{\parskip=0pt\parindent=0pt\par\vskip #1\noindent}}

\newcommand{\hscodestyle}{}

\newcommand{\sethscode}[1]%
  {\expandafter\let\expandafter\hscode\csname #1\endcsname
   \expandafter\let\expandafter\endhscode\csname end#1\endcsname}

  {\par\noindent
   \advance\leftskip\mathindent
   \hscodestyle
   \let\\=\@normalcr
   \let\hspre\(\let\hspost\)%
   \pboxed}%
  {\endpboxed\)%
   \par\noindent
   \ignorespacesafterend}

  {\hsnewpar\abovedisplayskip
   \advance\leftskip\mathindent
   \hscodestyle
   \let\hspre\(\let\hspost\)%
   \pboxed}%
  {\endpboxed%
   \hsnewpar\belowdisplayskip
   \ignorespacesafterend}

  {\hsnewpar\abovedisplayskip
   \advance\leftskip\mathindent
   \hscodestyle
   \let\\=\@normalcr
   \(\pboxed}%
  {\endpboxed\)%
   \hsnewpar\belowdisplayskip
   \ignorespacesafterend}

\newcommand{\plainhs}{\sethscode{plainhscode}}

\plainhs

  {\hsnewpar\abovedisplayskip
   \advance\leftskip\mathindent
   \hscodestyle
   \let\\=\@normalcr
   \(\parray}%
  {\endparray\)%
   \hsnewpar\belowdisplayskip
   \ignorespacesafterend}

  {\parray}{\endparray}

  {\(\parray}{\endparray\)}

\def\codeframewidth{\arrayrulewidth}
\RequirePackage{calc}

  {\parskip=\abovedisplayskip\par\noindent
   \hscodestyle
   \arrayrulewidth=\codeframewidth
   \tabular{@{}|p{\linewidth-2\arraycolsep-2\arrayrulewidth-2pt}|@{}}%
   \hline\framedhslinecorrect\\{-1.5ex}%
   \let\endoflinesave=\\
   \let\\=\@normalcr
   \(\pboxed}%
  {\endpboxed\)%
   \framedhslinecorrect\endoflinesave{.5ex}\hline
   \endtabular
   \parskip=\belowdisplayskip\par\noindent
   \ignorespacesafterend}

\newcommand{\framedhslinecorrect}[2]%
  {#1[#2]}

  {\(\def\column##1##2{}%
   \let\>\undefined\let\<\undefined\let\\\undefined
   \newcommand\>[1][]{}\newcommand\<[1][]{}\newcommand\\[1][]{}%
   \def\fromto##1##2##3{##3}%
   }{\) }%

  {\let\orighscode=\hscode
   \let\origendhscode=\endhscode
   \def\endhscode{\def\hscode{\endgroup\def\@currenvir{hscode}\\}\begingroup}
   \orighscode\def\hscode{\endgroup\def\@currenvir{hscode}}}%
  {\origendhscode
   \global\let\hscode=\orighscode
   \global\let\endhscode=\origendhscode}%

\makeatother
\EndFmtInput
\newcommand{\chooseFigPos}[1]{%
 \def\CFPPos{}
 \def\CFPTail{{figure}[\CFPPos]}
 \ifx\CFPPos\empty
  \begin{figure}[htbp]%
 \else\expandafter\begin\CFPTail\fi
}

\newenvironment{CenterFigure}[3]{%
  \def\EHCFigPos{\chooseFigPos{#1}}
  \def\EHCFigCap{#2}
  \def\EHCFigLab{#3}
  \EHCFigPos
  \begin{center}
}{%
  \end{center}
  \caption{\EHCFigCap}
  \label{\EHCFigLab}
  \end{figure}
}

\newenvironment{TabularCenterFigure}[4]{%
  \begin{CenterFigure}{#1}{#2}{#3}
  \begin{tabular}{#4}
}{%
  \end{tabular}
  \end{CenterFigure}
}

\newcommand{\rulerCmdDef}[1]{\expandafter\def\csname #1\endcsname}

\newcommand{\rulerCmdUse}[1]{\csname #1\endcsname}

\newcommand{\ruleName}[1]{\textsc{#1}}

\newenvironment{rulerRulesetFigure}[4]{%
  \def\RLRVersion{#4}
  \def\RLRInfo{\ifx\RLRVersion\empty #2\else #2\fi}
  \begin{TabularCenterFigure}{h}{\RLRInfo}{#3}{p{0.99\linewidth}}
  \small
  \begin{center}
  \fbox{#1}
  \\
}{%
  \par
  \csname RulesFigureBEndHook\endcsname
  \end{center}
  \end{TabularCenterFigure}
}

\newcommand{\rulerRuleHorNoView}[4]{%
\begingroup
\renewcommand{\arraystretch}{1}%
\ensuremath{%
\begin{array}{@{}l@{}}
\mbox{%
\ensuremath{%
  \frac{%
    \begin{array}{@{}c@{}}%
    \ \\
    #3
    \end{array}%
  }{%
    \begin{array}{@{}c@{}}%
    #4
    \end{array}%
  }%
 }%
\,
 {\footnotesize(\ruleName{#1})}
}
\\
\end{array}
}
\endgroup
}

\rulerCmdDef{typesystem.B.exp.scheme}{%
\ensuremath{\ensuremath{\Gamma\vdash } ^{ \ensuremath{e}} \ensuremath{e\mathbin{:}\sigma}}
}
\rulerCmdDef{typesystem.B.exp.exists.pack}{%
\rulerRuleHorNoView{pack}{B}
{ \ensuremath{\Gamma\vdash } ^{ \ensuremath{e}} \ensuremath{e\mathbin{:}[\mskip1.5mu \alpha\mapsto \sigma'\mskip1.5mu]\;\sigma} }
{ \ensuremath{\Gamma\vdash } ^{ \ensuremath{e}} \ensuremath{\lessdot(\alpha\mathrel{=}\sigma'), e\gtrdot\mathbin{:}\lessdot(\alpha\mathrel{=}\sigma'), \sigma\gtrdot} }
}

\rulerCmdDef{typesystem.B.exp.exists.eintro}{%
\rulerRuleHorNoView{E.I}{B}
{ \ensuremath{\Gamma\vdash } ^{ \ensuremath{e}} \ensuremath{e\mathbin{:}[\mskip1.5mu \alpha\mapsto \sigma\mskip1.5mu]\;\sigma'\to \mathbin{...}\to \lessdot(\alpha\mathrel{=}\sigma), \sigma''\gtrdot}
 }
{ \ensuremath{\Gamma\vdash } ^{ \ensuremath{e}} \ensuremath{{\bar{\exists}}\;\alpha\;.\;e\mathbin{:}{\bar{\exists}}\;\alpha\;.\;\sigma'\to \mathbin{...}\to \sigma''} }
}

\rulerCmdDef{typesystem.B.exp.exists.eelim}{%
\rulerRuleHorNoView{E.E}{B}
{ \ensuremath{\Gamma\vdash } ^{ \ensuremath{e}} \ensuremath{e\mathbin{:}{\bar{\exists}}\;\alpha\;.\;\sigma'\to \mathbin{...}\to \sigma''} }
{ \ensuremath{\Gamma,\nu\vdash } ^{ \ensuremath{e}} \ensuremath{e\;[\mskip1.5mu \nu\mskip1.5mu]\mathbin{:}[\mskip1.5mu \alpha\mapsto \nu\mskip1.5mu]\;\sigma'\to \mathbin{...}\to \lessdot(\alpha\mathrel{=}\nu), \sigma''\gtrdot}}
}

\rulerCmdDef{typesystem.B.exp.case}{%
\rulerRuleHorNoView{Choice}{B}
{ \ensuremath{\Gamma\vdash } ^{ \ensuremath{e}} \ensuremath{e_{\Varid{i}}\mathbin{:}\sigma}}
{ \ensuremath{\Gamma\vdash } ^{ \ensuremath{e}} \ensuremath{(e_{\mathrm{1}}\;\talloblong\mathbin{...}\talloblong\;e_{\Varid{n}})\mathbin{:}\sigma} }
}

\rulerCmdDef{typesystem.B.exp.let}{%
\rulerRuleHorNoView{Letrec}{B}
{ \ensuremath{\Lambda\vdash } ^{ \ensuremath{\Varid{p}}} \ensuremath{\Varid{p}\mathbin{:}\sigma}
\\\ensuremath{\Gamma\;\Lambda\vdash } ^{ \ensuremath{e}} \ensuremath{e_{\mathrm{1}}\mathbin{:}\sigma}
\\\ensuremath{\Gamma\;\Lambda\vdash } ^{ \ensuremath{e}} \ensuremath{e_2\mathbin{:}\sigma'}
}
{ \ensuremath{\Gamma\vdash } ^{ \ensuremath{e}} \ensuremath{\mathbf{let}\;\Varid{p}\mathrel{=}e_{\mathrm{1}}\;\mathbf{in}\;e_2\mathbin{:}\sigma'} }
}

\rulerCmdDef{typesystem.B.p.scheme}{%
\ensuremath{\ensuremath{\Lambda\vdash } ^{ \ensuremath{\Varid{p}}} \ensuremath{\Varid{p}\mathbin{:}\sigma}}
}

\rulerCmdDef{typesystem.B.p.var}{%
\rulerRuleHorNoView{varpat}{B}
{  }
{ \ensuremath{\Lambda,v\to \sigma\vdash } ^{ \ensuremath{\Varid{p}}} \ensuremath{v\mathbin{:}\sigma}}
}

\rulerCmdDef{typesystem.B.p.unpack}{%
\rulerRuleHorNoView{unpackpat}{B}
{  \ensuremath{\nu\;\notin\;\Lambda}
\\ \ensuremath{\Lambda,\nu\vdash } ^{ \ensuremath{\Varid{p}}} \ensuremath{\Varid{p}\mathbin{:}[\mskip1.5mu \alpha\mapsto \nu\mskip1.5mu]\;\sigma} }
{  \ensuremath{\Lambda,\nu\vdash } ^{ \ensuremath{\Varid{p}}} \ensuremath{\lessdot\nu, \Varid{p}\gtrdot\mathbin{:}\lessdot(\alpha\mathrel{=}\nu), \sigma\gtrdot}}
}

\rulerCmdDef{typesystem.B.exp}{%
\begin{rulerRulesetFigure}{\rulerCmdUse{typesystem.B.exp.scheme}}{Expression}{typesystem.B.exp}{B}
\rulerCmdUse{typesystem.B.exp.exists.pack}
\hspace{1ex}
\rulerCmdUse{typesystem.B.exp.exists.eintro}
\hspace{1ex}
\rulerCmdUse{typesystem.B.exp.exists.eelim}
\hspace{1ex}
\rulerCmdUse{typesystem.B.exp.case}
\hspace{1ex}
\rulerCmdUse{typesystem.B.exp.let}
\end{rulerRulesetFigure}
}

\rulerCmdDef{typesystem.B.p}{%
\begin{rulerRulesetFigure}{\rulerCmdUse{typesystem.B.p.scheme}}{Pattern}{typesystem.B.p}{B}
\rulerCmdUse{typesystem.B.p.var}
\hspace{1ex}
\rulerCmdUse{typesystem.B.p.unpack}
\end{rulerRulesetFigure}
}

\rulerCmdDef{typesystem.B.exp.varonly}{%
\begin{rulerRulesetFigure}{\rulerCmdUse{typesystem.B.exp.scheme}}{variable rule}{typesystem.B.exp}{B}
\rulerCmdUse{typesystem.B.exp.var}
\end{rulerRulesetFigure}
}

\usepackage{color}
\usepackage{graphicx}
\usepackage{mathpartir}

\begin{document}

\TRtitlepage
  {{A Lazy Language Needs a Lazy Type System: Introducing Polymorphic Contexts}}  
  {S. Doaitse Swierstra\\ Marcos Viera\\ Atze Dijkstra}                  
  {UU-CS-2016-012}                                         
  {Dec 2016}                                             

\title{A Lazy Language Needs a Lazy Type System\\ Introduing
  Polymorphic Contexts}
\author{S. Doaitse Swierstra and Marcos Viera and Atze Dijkstra}
\maketitle

\begin{abstract} 
Most type systems that support
polymorphic functions are based on a version of System-F. We argue
that this limits useful programming paradigms  for languages with lazy evaluation. 
We motivate an extension of System-F alleviating this limitation. 

First, using a sequence of examples, we show that for lazily
evaluated languages current type systems may force one  to write a
program in an unnatural way; we in particular argue that in such
languages the relationship between polymorphic and
existential types can be made more systematic by allowing to pass back
(part of) an existential
result of a function call as an argument to the the function call that produced
that value.
 
After presenting our extension to System-F we show how we can implement the strict-state thread
monad \ensuremath{\Conid{ST}} by using a returned existential type in
specialising the polymorphic function which returns that type. 
Currently this monad is  built-in into the runtime system of
GHC and as such has become part of the language.

Our proposed language extension, i.e. the introduction of
\emph{polymorphic contexts}, reverses the relationship between the context of a
function call and the called function with respect to  where it is decided with which type to
instantiate a type variable.
\end{abstract}
\tableofcontents

\section{Introduction}

In a strict language a value of type 
\begin{hscode}\SaveRestoreHook
\column{B}{@{}>{\hspre}l<{\hspost}@{}}%
\column{3}{@{}>{\hspre}l<{\hspost}@{}}%
\column{E}{@{}>{\hspre}l<{\hspost}@{}}%
\>[3]{}\exists\;\Varid{x}\;.\;\Varid{x}\to \Conid{Int}\to (\Varid{x},\Conid{Int}){}\<[E]%
\ColumnHook
\end{hscode}\resethooks
does not make sense; 
such functions cannot be called  since there is no way we can
provide the function with a useful first argument since the type \ensuremath{\Varid{x}}
is unknown.
In languages with lazy evaluation and \ensuremath{\mathbf{letrec}} bindings we can however
such an argument by passing part of the computed result:
\begin{hscode}\SaveRestoreHook
\column{B}{@{}>{\hspre}l<{\hspost}@{}}%
\column{6}{@{}>{\hspre}l<{\hspost}@{}}%
\column{11}{@{}>{\hspre}l<{\hspost}@{}}%
\column{E}{@{}>{\hspre}l<{\hspost}@{}}%
\>[B]{}\mathbf{let}\;{}\<[6]%
\>[6]{}\Varid{f}\mathbin{::}\exists\;\Varid{x}\;.\;\Varid{x}\to \Conid{Int}\to (\Varid{x},\Conid{Int})\mathrel{=}\mathbin{...}{}\<[E]%
\\
\>[B]{}\mathbf{in}\;{}\<[6]%
\>[6]{}\mathbf{let}\;{}\<[11]%
\>[11]{}(\Varid{x},\Varid{v})\mathrel{=}\Varid{f}\;\Varid{x}\;\mathrm{3}{}\<[E]%
\\
\>[6]{}\mathbf{in}\;{}\<[11]%
\>[11]{}\Varid{v}{}\<[E]%
\ColumnHook
\end{hscode}\resethooks

We claim that conventional System-F based type systems exclude
some useful programming paradigms and thus we propose a small
extension to System-F.  The larger part of this paper consists of two
examples in which we show how to put our extension to good use. 
We show in Section \ref{origin} how
the current way of dealing with existential types in Haskell (GHC)
forces undesirable strictness on our programs and can make our
programs unnecessarily complicated. 
Next we discuss our type system in Section \ref{polycontexts}. 
We then proceed by showing in Section \ref{stmonad} how we can encode the
\ensuremath{\Conid{ST}}-monad, so there is no longer the need to have it built into the
language. We finish with some possible extensions, discussion and conclusions.

\section{Being less strict}\label{origin} 
In order to explain 
what kind of programs we should like to write, we start out with the 
\ensuremath{\Varid{repmin}} problem \cite{springerlink:10.1007/BF00264249}. We start with
a lazy version and convert that into a strict version that performs  the same
number of pattern matches. Next we write a similar version of
an identity function. When we convert that function back to a lazy version,
similar to the first version of \ensuremath{\Varid{repmin}} we run into a typing problem.  

\subsection{\ensuremath{\Varid{repmin}}} 

The challenge is
to write a function \ensuremath{\Varid{repmin}\mathbin{::}\Conid{Tree}\to \Conid{Tree}} which returns a binary tree
with the same shape as the argument tree, but with the leave values 
replaced by the minimum of the original leaf values. A
straightforward solution, which also can be seen as a specification of
the problem, is given in Figure~\ref{fig:repminstrict}. Note that in
this solution no use of lazy evaluation is made and that each node of
the tree is inspected twice in a pattern match during the computation. 

\begin{figure}
\begin{hscode}\SaveRestoreHook
\column{B}{@{}>{\hspre}l<{\hspost}@{}}%
\column{12}{@{}>{\hspre}c<{\hspost}@{}}%
\column{12E}{@{}l@{}}%
\column{13}{@{}>{\hspre}l<{\hspost}@{}}%
\column{15}{@{}>{\hspre}l<{\hspost}@{}}%
\column{18}{@{}>{\hspre}l<{\hspost}@{}}%
\column{20}{@{}>{\hspre}l<{\hspost}@{}}%
\column{23}{@{}>{\hspre}l<{\hspost}@{}}%
\column{E}{@{}>{\hspre}l<{\hspost}@{}}%
\>[B]{}\mathbf{data}\;\Conid{Tree}{}\<[12]%
\>[12]{}\mathrel{=}{}\<[12E]%
\>[15]{}\Conid{Leaf}\;\Conid{Int}{}\<[E]%
\\
\>[12]{}\mid {}\<[12E]%
\>[15]{}\Conid{Bin}\;\Conid{Tree}\;\Conid{Tree}{}\<[E]%
\\[\blanklineskip]%
\>[B]{}\Varid{repmin}\;\Varid{t}\mathrel{=}{}\<[13]%
\>[13]{}\mathbf{let}\;{}\<[18]%
\>[18]{}\Varid{m}\mathrel{=}\Varid{minval}\;\Varid{t}\;\mathbf{in}\;\Varid{replace}\;\Varid{t}\;\Varid{m}{}\<[E]%
\\[\blanklineskip]%
\>[B]{}\Varid{minval}\mathbin{::}\Conid{Tree}\to \Conid{Int}{}\<[E]%
\\
\>[B]{}\Varid{minval}\;(\Conid{Leaf}\;\Varid{v}){}\<[23]%
\>[23]{}\mathrel{=}\Varid{v}{}\<[E]%
\\
\>[B]{}\Varid{minval}\;(\Conid{Bin}\;\Varid{l}\;\Varid{r}){}\<[23]%
\>[23]{}\mathrel{=}\Varid{minval}\;\Varid{l}\mathbin{`\Varid{min}`}\Varid{minval}\;\Varid{r}{}\<[E]%
\\[\blanklineskip]%
\>[B]{}\Varid{replace}\mathbin{::}\Conid{Tree}\to \Conid{Int}\to \Conid{Tree}{}\<[E]%
\\
\>[B]{}\Varid{replace}\;(\Conid{Leaf}\;\anonymous )\;{}\<[20]%
\>[20]{}\Varid{m}{}\<[23]%
\>[23]{}\mathrel{=}\Conid{Leaf}\;\Varid{m}{}\<[E]%
\\
\>[B]{}\Varid{replace}\;(\Conid{Bin}\;\Varid{l}\;\Varid{r})\;{}\<[20]%
\>[20]{}\Varid{m}{}\<[23]%
\>[23]{}\mathrel{=}\Varid{replace}\;\Varid{l}\;\Varid{m}\mathbin{`\Conid{Bin}`}\Varid{replace}\;\Varid{r}\;\Varid{m}{}\<[E]%
\ColumnHook
\end{hscode}\resethooks
\caption{\ensuremath{\Varid{repmin}} in a strict language}\label{fig:repminstrict}
\end{figure}

The reason that this problem has drawn a lot of
attention is that, provided the programming language supports lazy evaluation (call by need), the
result can be computed by inspecting each constructor of the argument
tree only once (Figure~\ref{fig:repmin}): in the function \ensuremath{\Varid{repmin'}} we
tuple the computation of the minimal leaf value with the construction
of the resulting tree. The latter uses the computed minimal
value which is passed as the parameter \ensuremath{\Varid{m}}. In the top function \ensuremath{\Varid{repmin}} the
minimal value computed by \ensuremath{\Varid{repmin'}} is {\em passed back} as argument
to the same call of \ensuremath{\Varid{repmin'}} as the value to be bound to argument \ensuremath{\Varid{m}}. Such programs are referred to as \emph{circular programs}. 

\begin{figure}
\begin{hscode}\SaveRestoreHook
\column{B}{@{}>{\hspre}l<{\hspost}@{}}%
\column{13}{@{}>{\hspre}l<{\hspost}@{}}%
\column{18}{@{}>{\hspre}l<{\hspost}@{}}%
\column{20}{@{}>{\hspre}l<{\hspost}@{}}%
\column{23}{@{}>{\hspre}c<{\hspost}@{}}%
\column{23E}{@{}l@{}}%
\column{26}{@{}>{\hspre}l<{\hspost}@{}}%
\column{33}{@{}>{\hspre}l<{\hspost}@{}}%
\column{39}{@{}>{\hspre}l<{\hspost}@{}}%
\column{44}{@{}>{\hspre}l<{\hspost}@{}}%
\column{E}{@{}>{\hspre}l<{\hspost}@{}}%
\>[B]{}\mathbf{data}\;\Conid{Tree}\mathrel{=}\Conid{Leaf}\;\Conid{Int}\mid \Conid{Bin}\;\Conid{Tree}\;\Conid{Tree}{}\<[E]%
\\[\blanklineskip]%
\>[B]{}\Varid{repmin}\mathbin{::}\Conid{Tree}\to \Conid{Tree}{}\<[E]%
\\
\>[B]{}\Varid{repmin}\;\Varid{t}\mathrel{=}{}\<[13]%
\>[13]{}\mathbf{let}\;{}\<[18]%
\>[18]{}(\Varid{m},\Varid{r})\mathrel{=}\Varid{repmin'}\;\Varid{t}\;\Varid{m}\;\mathbf{in}\;\Varid{r}{}\<[E]%
\\[\blanklineskip]%
\>[B]{}\Varid{repmin'}\mathbin{::}\Conid{Tree}\to \Conid{Int}\to (\Conid{Int},\Conid{Tree}){}\<[E]%
\\
\>[B]{}\Varid{repmin'}\;(\Conid{Leaf}\;\Varid{v})\;{}\<[20]%
\>[20]{}\Varid{m}{}\<[23]%
\>[23]{}\mathrel{=}{}\<[23E]%
\>[26]{}(\Varid{v},\Conid{Leaf}\;\Varid{m}){}\<[E]%
\\
\>[B]{}\Varid{repmin'}\;(\Conid{Bin}\;\Varid{l}\;\Varid{r})\;{}\<[20]%
\>[20]{}\Varid{m}{}\<[23]%
\>[23]{}\mathrel{=}{}\<[23E]%
\>[26]{}(\Varid{ml}\mathbin{`\Varid{min}`}\Varid{mr},\Varid{tl}\mathbin{`\Conid{Bin}`}\Varid{tr}){}\<[E]%
\\
\>[26]{}\mathbf{where}\;{}\<[33]%
\>[33]{}(\Varid{ml},{}\<[39]%
\>[39]{}\Varid{tl}){}\<[44]%
\>[44]{}\mathrel{=}\Varid{repmin'}\;\Varid{l}\;\Varid{m}{}\<[E]%
\\
\>[33]{}(\Varid{mr},{}\<[39]%
\>[39]{}\Varid{tr}){}\<[44]%
\>[44]{}\mathrel{=}\Varid{repmin'}\;\Varid{r}\;\Varid{m}{}\<[E]%
\ColumnHook
\end{hscode}\resethooks
\caption{\ensuremath{\Varid{repmin}} in a lazy language}\label{fig:repmin}
\end{figure}

If we perform a global flow analysis of the program, inspired by
analyses from the attribute-grammar world
\cite{kastens80order-attrgam,Engelfriet1982283,vanBinsbergen:2015:LOA:2678015.2682543},
we discover that no information flows from the \ensuremath{\Conid{Int}} parameter to the
\ensuremath{\Conid{Int}} part of the result.  This implies that we can replace the type
\ensuremath{\Conid{Tree}\to \Conid{Int}\to (\Conid{Int},\Conid{Tree})} with the type \ensuremath{\Conid{Tree}\to (\Conid{Int},\Conid{Int}\to \Conid{Tree})},
provided the program is adapted accordingly. Figure
\ref{repmin2} shows the result of this transformation.
The \ensuremath{\Conid{Int}} part of the result again 
contains the computed minimal value and the  \ensuremath{\Conid{Int}\to \Conid{Tree}}
part is a function that constructs
the sought tree from the passed minimal value; we have
``remembered'' the shape of the argument tree in that function.

\begin{figure}
\begin{hscode}\SaveRestoreHook
\column{B}{@{}>{\hspre}l<{\hspost}@{}}%
\column{14}{@{}>{\hspre}l<{\hspost}@{}}%
\column{21}{@{}>{\hspre}c<{\hspost}@{}}%
\column{21E}{@{}l@{}}%
\column{24}{@{}>{\hspre}l<{\hspost}@{}}%
\column{31}{@{}>{\hspre}l<{\hspost}@{}}%
\column{42}{@{}>{\hspre}l<{\hspost}@{}}%
\column{E}{@{}>{\hspre}l<{\hspost}@{}}%
\>[B]{}\Varid{repmin2}\mathbin{::}\Conid{Tree}\to \Conid{Tree}{}\<[E]%
\\
\>[B]{}\Varid{repmin2}\;\Varid{t}\mathrel{=}{}\<[14]%
\>[14]{}\mathbf{let}\;(\Varid{m},\Varid{reconstruct})\mathrel{=}\Varid{repmin2'}\;\Varid{t}{}\<[E]%
\\
\>[14]{}\mathbf{in}\;\Varid{reconstruct}\;\Varid{m}{}\<[E]%
\\[\blanklineskip]%
\>[B]{}\Varid{repmin2'}\mathbin{::}\Conid{Tree}\to (\Conid{Int},\Conid{Int}\to \Conid{Tree}){}\<[E]%
\\
\>[B]{}\Varid{repmin2'}\;(\Conid{Leaf}\;\Varid{v}){}\<[21]%
\>[21]{}\mathrel{=}{}\<[21E]%
\>[24]{}(\Varid{v},\Conid{Leaf}){}\<[E]%
\\
\>[B]{}\Varid{repmin2'}\;(\Conid{Bin}\;\Varid{l}\;\Varid{r}){}\<[21]%
\>[21]{}\mathrel{=}{}\<[21E]%
\>[24]{}(\Varid{ml}\mathbin{`\Varid{min}`}\Varid{mr}{}\<[E]%
\\
\>[24]{},\lambda \Varid{m}\to \Varid{tfl}\;\Varid{m}\mathbin{`\Conid{Bin}`}\Varid{tfr}\;\Varid{m}){}\<[E]%
\\
\>[24]{}\mathbf{where}\;{}\<[31]%
\>[31]{}(\Varid{ml},\Varid{tfl}){}\<[42]%
\>[42]{}\mathrel{=}\Varid{repmin2'}\;\Varid{l}{}\<[E]%
\\
\>[31]{}(\Varid{mr},\Varid{tfr}){}\<[42]%
\>[42]{}\mathrel{=}\Varid{repmin2'}\;\Varid{r}{}\<[E]%
\ColumnHook
\end{hscode}\resethooks
\caption{strict version of \ensuremath{\Varid{repmin2}}}\label{repmin2}
\end{figure} 

This code maintains the characterizing property of our lazy \ensuremath{\Varid{repmin}}
solution: each constructor of the tree is only inspected once in a
pattern match; the order in which values are to be evaluated however
has been made more explicit (although lazy evaluation still evaluates
them in the same order!).  Note that the first version of \ensuremath{\Varid{repmin}}
depends essentially on lazy evaluation (the \ensuremath{\mathbf{let}} actually is a
\ensuremath{\mathbf{letrec}} in Haskell), whereas \ensuremath{\Varid{repmin2}}, despite being here written in
Haskell, could straightforwardly be transcribed into a strict language
like ML.

\subsection{\ensuremath{\Varid{idTree}}}
The next step in introducing our problem is that instead of writing a
\ensuremath{\Varid{repmin}} function we want to write an (admittedly overly complex and
utterly useless) identity function of type \ensuremath{\Conid{Tree}\to \Conid{Tree}}.  Where
\ensuremath{\Varid{repmin2}} computed an intermediate representation holding the minimal
value of the leaves tupled with a function which remembered the
\emph{shape} of the tree, our first identity function \ensuremath{\Varid{idTree2}}
(Figure~\ref{fig:idTree2}) uses a similar intermediate structure
which contains all the values stored in the leaves of the
original tree in a nested Cartesian product tupled with a
tree-reconstruction function.  The latter function, as before, has
remembered the shape of the argument tree and, once provided with the
leaf values that were harvested from that tree, reconstructs that very
tree. Since, in contrast to the \ensuremath{\Varid{repmin2}} function, \emph{the type of
  the intermediate result depends on the shape of the tree}, we have
introduced an existential type \ensuremath{\Varid{vs}} in this
representation.

To make explicit what is going on we introduce
some notation to make the places where existential values are
constructed and deconstructed explicit using the
\emph{pack/unpack} paradigm 
\cite{Mitchell:1988:ATE:44501.45065}.
In a \emph{packing} expression \ensuremath{\lessdot\Varid{t}, \Varid{e}\gtrdot}
the \ensuremath{\Varid{t}} denotes the existential type and the \ensuremath{\Varid{e}} a value of a
type in which this type may occur. The \emph{unpack} function is implicitly
called by using
pattern matching (following Pierce \cite{Pierce}); binding to  a \ensuremath{\lessdot\Varid{t}_{\Varid{v}}, \Varid{v}\gtrdot} pattern makes that the a
freshly new  type
constant is bound to the type variable \ensuremath{\Varid{t}_{\Varid{v}}} and the value part to the
variable \ensuremath{\Varid{v}}.

When pairing the two parts of the result in the branches of
\ensuremath{\Varid{idTree2'}} we thus
hide whether we combine an \ensuremath{\Conid{Int}} value with a function of type \ensuremath{\Conid{Int}\to \Conid{Tree}}  as in the first alternative of \ensuremath{\Varid{idTree2'}} or a pair of existentially 
typed values returned by the recursive function calls with a function taking such a pair as in the second alternative.
When we unpack the packed value in \ensuremath{\Varid{idTree2}} using pattern matching we can however be sure that it is safe to apply 
the function to its accompanying value, because this was the case when we packed them together.

\begin{figure} 
\begin{hscode}\SaveRestoreHook
\column{B}{@{}>{\hspre}l<{\hspost}@{}}%
\column{4}{@{}>{\hspre}c<{\hspost}@{}}%
\column{4E}{@{}l@{}}%
\column{7}{@{}>{\hspre}l<{\hspost}@{}}%
\column{12}{@{}>{\hspre}l<{\hspost}@{}}%
\column{14}{@{}>{\hspre}l<{\hspost}@{}}%
\column{18}{@{}>{\hspre}l<{\hspost}@{}}%
\column{20}{@{}>{\hspre}c<{\hspost}@{}}%
\column{20E}{@{}l@{}}%
\column{23}{@{}>{\hspre}l<{\hspost}@{}}%
\column{32}{@{}>{\hspre}l<{\hspost}@{}}%
\column{38}{@{}>{\hspre}l<{\hspost}@{}}%
\column{39}{@{}>{\hspre}l<{\hspost}@{}}%
\column{49}{@{}>{\hspre}l<{\hspost}@{}}%
\column{E}{@{}>{\hspre}l<{\hspost}@{}}%
\>[B]{}\Varid{idTree2}\mathbin{::}\Conid{Tree}\to \Conid{Tree}{}\<[E]%
\\
\>[B]{}\Varid{idTree2}\;\Varid{t}\mathrel{=}{}\<[14]%
\>[14]{}\mathbf{let}\;\lessdot\Varid{t}_{\Varid{vs}}, (\Varid{vs},\Varid{reconstruct})\gtrdot\mathrel{=}{}\<[49]%
\>[49]{}\Varid{idTree2'}\;\Varid{t}{}\<[E]%
\\
\>[14]{}\mathbf{in}\;{}\<[18]%
\>[18]{}\Varid{reconstruct}\;\Varid{vs}{}\<[E]%
\\[\blanklineskip]%
\>[B]{}\Varid{idTree2'}\mathbin{::}\Conid{Tree}\to \exists\;\Varid{vs}\;.\;(\Varid{vs},{}\<[39]%
\>[39]{}(\Varid{vs}\to \Conid{Tree})){}\<[E]%
\\
\>[B]{}\Varid{idTree2'}\;(\Conid{Leaf}\;\Varid{v}){}\<[20]%
\>[20]{}\mathrel{=}{}\<[20E]%
\>[23]{}\lessdot\Conid{Int}, (\Varid{v},\Conid{Leaf})\gtrdot{}\<[E]%
\\
\>[B]{}\Varid{idTree2'}\;(\Conid{Bin}\;\Varid{l}\;\Varid{r}){}\<[E]%
\\
\>[B]{}\hsindent{4}{}\<[4]%
\>[4]{}\mathrel{=}{}\<[4E]%
\>[7]{}\lessdot(\Varid{t}_{\Varid{vsl}},\Varid{t}_{\Varid{vsr}}), {}\<[E]%
\\
\>[7]{}\hsindent{5}{}\<[12]%
\>[12]{}((\Varid{vsl},\Varid{vsr}),(\lambda (\Varid{vsl},\Varid{vsr})\to \Varid{tfl}\;\Varid{vsl}\mathbin{`\Conid{Bin}`}\Varid{tfr}\;\Varid{vsr}))\gtrdot{}\<[E]%
\\
\>[7]{}\mathbf{where}\;{}\<[14]%
\>[14]{}\lessdot\Varid{t}_{\Varid{vsl}}, (\Varid{vsl},{}\<[32]%
\>[32]{}\Varid{tfl})\gtrdot{}\<[38]%
\>[38]{}\mathrel{=}\Varid{idTree2'}\;\Varid{l}{}\<[E]%
\\
\>[14]{}\lessdot\Varid{t}_{\Varid{vsr}}, (\Varid{vsr},{}\<[32]%
\>[32]{}\Varid{tfr})\gtrdot{}\<[38]%
\>[38]{}\mathrel{=}\Varid{idTree2'}\;\Varid{r}{}\<[E]%
\ColumnHook
\end{hscode}\resethooks
\caption{\ensuremath{\Varid{idTree2}}}\label{fig:idTree2}
\end{figure}

Now suppose we have a language with lazy
evaluation and that we prefer the lazy version of \ensuremath{\Varid{repmin}} over \ensuremath{\Varid{repmin2}},  and thus we
set out to define a similar \ensuremath{\Varid{idTree}}  (Figure \ref{idTree}), in which
we do not use an intermediate representation.

\begin{figure}
\begin{hscode}\SaveRestoreHook
\column{B}{@{}>{\hspre}l<{\hspost}@{}}%
\column{10}{@{}>{\hspre}l<{\hspost}@{}}%
\column{13}{@{}>{\hspre}l<{\hspost}@{}}%
\column{14}{@{}>{\hspre}l<{\hspost}@{}}%
\column{18}{@{}>{\hspre}l<{\hspost}@{}}%
\column{21}{@{}>{\hspre}l<{\hspost}@{}}%
\column{24}{@{}>{\hspre}l<{\hspost}@{}}%
\column{27}{@{}>{\hspre}l<{\hspost}@{}}%
\column{32}{@{}>{\hspre}c<{\hspost}@{}}%
\column{32E}{@{}l@{}}%
\column{35}{@{}>{\hspre}l<{\hspost}@{}}%
\column{46}{@{}>{\hspre}l<{\hspost}@{}}%
\column{E}{@{}>{\hspre}l<{\hspost}@{}}%
\>[B]{}\Varid{idTree}\mathbin{::}\Conid{Tree}\to \Conid{Tree}{}\<[E]%
\\
\>[B]{}\Varid{idTree}\;\Varid{t}\mathrel{=}{}\<[13]%
\>[13]{}\mathbf{let}\;{}\<[18]%
\>[18]{}(\Varid{vs},\Varid{r})\mathrel{=}\Varid{idTree'}\;\Varid{t}\;\Varid{vs}\;\mathbf{in}\;\Varid{r}{}\<[E]%
\\[\blanklineskip]%
\>[B]{}\Varid{idTree'}\;{}\<[10]%
\>[10]{}(\Conid{Leaf}\;\Varid{v})\;{}\<[24]%
\>[24]{}\Varid{w}{}\<[E]%
\\
\>[10]{}\mathrel{=}{}\<[13]%
\>[13]{}(\Varid{v},{}\<[27]%
\>[27]{}\Conid{Leaf}\;\Varid{w}){}\<[E]%
\\
\>[B]{}\Varid{idTree'}\;{}\<[10]%
\>[10]{}(\Conid{Bin}\;\Varid{l}\;\Varid{r})\; {}\<[24]%
\>[24]{}\mathord{\sim}(\Varid{vsl'},\Varid{vsr'}){}\<[E]%
\\
\>[10]{}\mathrel{=}{}\<[13]%
\>[13]{}((\Varid{vsl},\Varid{vsr}),{}\<[27]%
\>[27]{}\Varid{tl}\mathbin{`\Conid{Bin}`}\Varid{tr}){}\<[E]%
\\
\>[13]{}\hsindent{1}{}\<[14]%
\>[14]{}\mathbf{where}\;{}\<[21]%
\>[21]{}(\Varid{vsl},\Varid{tl}){}\<[32]%
\>[32]{}\mathrel{=}{}\<[32E]%
\>[35]{}\Varid{idTree'}\;\Varid{l}\;{}\<[46]%
\>[46]{}\Varid{vsl'}{}\<[E]%
\\
\>[21]{}(\Varid{vsr},\Varid{tr}){}\<[32]%
\>[32]{}\mathrel{=}{}\<[32E]%
\>[35]{}\Varid{idTree'}\;\Varid{r}\;{}\<[46]%
\>[46]{}\Varid{vsr'}{}\<[E]%
\ColumnHook
\end{hscode}\resethooks
\caption{A type incorrect \ensuremath{\Varid{idTree}}}\label{idTree}
\end{figure}

We now run into a problem, since this program is
type-incorrect and we cannot provide a type for \ensuremath{\Varid{idTree'}}.  In the
first alternative of \ensuremath{\Varid{idTree'}} the argument \ensuremath{\Varid{w}} is of type \ensuremath{\Conid{Int}},
whereas in the second alternative the type is a pair of values, and
these two types do not unify.  So, why is the first version of
\ensuremath{\Varid{repmin}} permitted, and is our corresponding version of \ensuremath{\Varid{idTree}} is
rejected? When checking the types at runtime we do
not run into problems.

It appears that we actually are in need of a dependent type; the type
of the returned structure containing the leaf values, and thus the
type of the second argument which holds the leaf values from which the
tree is to be reconstructed, entirely
depends on the shape of the argument tree. By rewriting the code, and
making use of some explicit lambda's we reach the solution given in Figure
\ref{fig:idTreeppghc}. 

\begin{figure}
\begin{hscode}\SaveRestoreHook
\column{B}{@{}>{\hspre}l<{\hspost}@{}}%
\column{4}{@{}>{\hspre}l<{\hspost}@{}}%
\column{9}{@{}>{\hspre}l<{\hspost}@{}}%
\column{13}{@{}>{\hspre}l<{\hspost}@{}}%
\column{15}{@{}>{\hspre}l<{\hspost}@{}}%
\column{18}{@{}>{\hspre}l<{\hspost}@{}}%
\column{20}{@{}>{\hspre}c<{\hspost}@{}}%
\column{20E}{@{}l@{}}%
\column{24}{@{}>{\hspre}l<{\hspost}@{}}%
\column{36}{@{}>{\hspre}l<{\hspost}@{}}%
\column{41}{@{}>{\hspre}l<{\hspost}@{}}%
\column{51}{@{}>{\hspre}l<{\hspost}@{}}%
\column{57}{@{}>{\hspre}l<{\hspost}@{}}%
\column{E}{@{}>{\hspre}l<{\hspost}@{}}%
\>[B]{}\Varid{idTree}\mathbin{::}\Conid{Tree}\to \Conid{Tree}{}\<[E]%
\\
\>[B]{}\Varid{idTree}\;\Varid{t}\mathrel{=}{}\<[13]%
\>[13]{}\mathbf{let}\;{}\<[18]%
\>[18]{}\lessdot\Varid{t}_{\Varid{vs}}, \Varid{f}\gtrdot\mathrel{=}\Varid{idTree'}\;\Varid{t}{}\<[E]%
\\
\>[18]{}(\Varid{vs},\Varid{r})\mathrel{=}\Varid{f}\;\Varid{vs}{}\<[E]%
\\
\>[13]{}\mathbf{in}\;\Varid{r}{}\<[E]%
\\[\blanklineskip]%
\>[B]{}\Varid{idTree'}\mathbin{::}\Conid{Tree}\to \exists\;\Varid{vs}\;.\;\Varid{vs}\to (\Varid{vs},\Conid{Tree}){}\<[E]%
\\
\>[B]{}\Varid{idTree'}\;(\Conid{Leaf}\;\Varid{v}){}\<[20]%
\>[20]{}\mathrel{=}{}\<[20E]%
\>[24]{}\lessdot\Conid{Int}, (\lambda \Varid{w}\to (\Varid{v},\Conid{Leaf}\;\Varid{w}))\gtrdot{}\<[E]%
\\
\>[B]{}\Varid{idTree'}\;(\Conid{Bin}\;\Varid{l}\;\Varid{r}){}\<[20]%
\>[20]{}\mathrel{=}{}\<[20E]%
\\
\>[B]{}\hsindent{4}{}\<[4]%
\>[4]{}\mathbf{let}\;{}\<[9]%
\>[9]{}\lessdot\Varid{t}_{\Varid{vsl}}, \Varid{fl}\gtrdot{}\<[24]%
\>[24]{}\mathrel{=}\Varid{idTree'}\;\Varid{l}{}\<[E]%
\\
\>[9]{}\lessdot\Varid{t}_{\Varid{vsr}}, \Varid{fr}\gtrdot{}\<[24]%
\>[24]{}\mathrel{=}\Varid{idTree'}\;\Varid{r}{}\<[E]%
\\
\>[B]{}\hsindent{4}{}\<[4]%
\>[4]{}\mathbf{in}\;{}\<[9]%
\>[9]{}\lessdot{}\<[15]%
\>[15]{}(\Varid{t}_{\Varid{vsl}},\Varid{t}_{\Varid{vsr}}), {}\<[E]%
\\
\>[15]{}(\lambda \mathord{\sim}(\Varid{vsl'},\Varid{vsr'})\to {}\<[36]%
\>[36]{}\mathbf{let}\;{}\<[41]%
\>[41]{}(\Varid{vsl},\Varid{tl}){}\<[51]%
\>[51]{}\mathrel{=}\Varid{fl}\;{}\<[57]%
\>[57]{}\Varid{vsl'}{}\<[E]%
\\
\>[41]{}(\Varid{vsr},\Varid{tr}){}\<[51]%
\>[51]{}\mathrel{=}\Varid{fr}\;{}\<[57]%
\>[57]{}\Varid{vsr'}{}\<[E]%
\\
\>[36]{}\mathbf{in}\;{}\<[41]%
\>[41]{}((\Varid{vsl},\Varid{vsr}),\Varid{tl}\mathbin{`\Conid{Bin}`}\Varid{tr}))\gtrdot{}\<[E]%
\ColumnHook
\end{hscode}\resethooks
\caption{\ensuremath{\Varid{idTree}}, computing the types first}\label{fig:idTreeppghc}
\end{figure}

Unfortunately this version cannot be transcribed into GHC while
keeping its semantics. One of GHC's design decisions has been to
forbid irrefutable patterns, and thus all pattern matching for
existential data types has to be strict using a \ensuremath{\mathbf{case}} construct, which excludes the use of
\ensuremath{\mathbf{let}}-based bindings with existential types as their right hand side.
In requiring strict pattern matching the ``computation'' of the
complete type is enforced before being able to pack it with
the existential construct, and we have thus silently changed the
semantics of our original \ensuremath{\Varid{idTree}}.  This is demonstrated by the
program in Figure \ref{toostrict}. The GHC solution will not terminate
since it enforecs a complete traversal of an infinite tree trying to construct an infinite
type representing the shape of the infinite tree argument.  In this
sense the GHC solution is not an honest realization of the identity
function.

\begin{figure}
\begin{hscode}\SaveRestoreHook
\column{B}{@{}>{\hspre}l<{\hspost}@{}}%
\column{9}{@{}>{\hspre}l<{\hspost}@{}}%
\column{16}{@{}>{\hspre}l<{\hspost}@{}}%
\column{E}{@{}>{\hspre}l<{\hspost}@{}}%
\>[B]{}\Varid{top}\;(\Conid{Bin}\;\anonymous \;\anonymous ){}\<[16]%
\>[16]{}\mathrel{=}\text{\tt \char34 Bin\char34}{}\<[E]%
\\
\>[B]{}\Varid{top}\;\Conid{Leaf}{}\<[16]%
\>[16]{}\mathrel{=}\text{\tt \char34 Leaf\char34}{}\<[E]%
\\
\>[B]{}\Varid{main}\mathrel{=}{}\<[9]%
\>[9]{}\mathbf{let}\;\Varid{inftree}\mathrel{=}\Conid{Bin}\;\Varid{infTree}\;\Varid{infTree}{}\<[E]%
\\
\>[9]{}\mathbf{in}\;\Varid{print}\;.\;\Varid{top}\;.\;\Varid{idTree}\mathbin{\$}\Varid{inftree}{}\<[E]%
\ColumnHook
\end{hscode}\resethooks
\caption{GHC  version is too strict}\label{toostrict}
\end{figure}

A more serious, methodological shortcoming, of this last version of
\ensuremath{\Varid{idTree}} is that we had to separate the computation of the type from the
computation of the values: first we inspect the tree, this gives us
the types and constructs the computations to be performed, and then we
perform the computations. So can we make Figure \ref{idTree} type check?

\subsection{\ensuremath{{\bar{\exists}}}}
One way of looking at the type-incorrect \ensuremath{\Varid{idTree}} is not to see the type
of the gathered leaves as something which is computed by
inspecting the parameter tree, but as something which is computed by
the function too, and returned as part of its result; the function then
expects a value of this returned type as a lazy evaluated argument.  
Intuitively the type of \ensuremath{\Varid{idTree'}} is something like:

\begin{hscode}\SaveRestoreHook
\column{B}{@{}>{\hspre}l<{\hspost}@{}}%
\column{E}{@{}>{\hspre}l<{\hspost}@{}}%
\>[B]{}\Varid{idTree'}\mathbin{::}\Conid{Tree}\to \Varid{t}_{\Varid{vs}}\to \exists\;\Varid{t}_{\Varid{vs}}\;.\;(\Varid{t}_{\Varid{vs}},\Conid{Tree}){}\<[E]%
\ColumnHook
\end{hscode}\resethooks

This notation reflects the operational idea that the function
\ensuremath{\Varid{idTree'}} not only returns a value of type \ensuremath{\Varid{t}_{\Varid{vs}}}, but also the type
\ensuremath{\Varid{t}_{\Varid{vs}}} itself. The notation is however a bit unconventional since the
\ensuremath{\Varid{t}_{\Varid{vs}}} in the argument position is also supposed to be introduced by
this \ensuremath{\exists} quantifier, which is unfortunately not reflected in the
notation at all. To overcome this problem we introduce a new
quantifier \ensuremath{{\bar{\exists}}}, which is to be interpreted as the specification
above; hence it introduces a type variable which is bound to a type computed as part of the
result of the function, but which scopes over (part of) the
signature too:

\begin{hscode}\SaveRestoreHook
\column{B}{@{}>{\hspre}l<{\hspost}@{}}%
\column{35}{@{}>{\hspre}l<{\hspost}@{}}%
\column{E}{@{}>{\hspre}l<{\hspost}@{}}%
\>[B]{}\Varid{idTree'}\mathbin{::}\Conid{Tree}\to {\bar{\exists}}\;\Varid{t}_{\Varid{vs}}\;.\;{}\<[35]%
\>[35]{}\Varid{t}_{\Varid{vs}}\to (\Varid{t}_{\Varid{vs}},\Conid{Tree}){}\<[E]%
\ColumnHook
\end{hscode}\resethooks

By extending the scope to an earlier position in the list of
arguments, we express that both the \ensuremath{\Varid{t}_{\Varid{vs}}} argument  and the 
\ensuremath{\Varid{t}_{\Varid{vs}}} part of the result are to be the same for each call to \ensuremath{\Varid{idTree'}}, but furthermore
opaque at the calling position.

\begin{figure}
\begin{hscode}\SaveRestoreHook
\column{B}{@{}>{\hspre}l<{\hspost}@{}}%
\column{5}{@{}>{\hspre}l<{\hspost}@{}}%
\column{8}{@{}>{\hspre}c<{\hspost}@{}}%
\column{8E}{@{}l@{}}%
\column{11}{@{}>{\hspre}l<{\hspost}@{}}%
\column{13}{@{}>{\hspre}l<{\hspost}@{}}%
\column{16}{@{}>{\hspre}l<{\hspost}@{}}%
\column{25}{@{}>{\hspre}l<{\hspost}@{}}%
\column{27}{@{}>{\hspre}c<{\hspost}@{}}%
\column{27E}{@{}l@{}}%
\column{39}{@{}>{\hspre}l<{\hspost}@{}}%
\column{E}{@{}>{\hspre}l<{\hspost}@{}}%
\>[B]{}\Varid{idTree}\;\Varid{t}\mathrel{=}{}\<[13]%
\>[13]{}\mathbf{let}\;\lessdot\Varid{t}_{\Varid{vs}}, (\Varid{vs},\Varid{r})\gtrdot\mathrel{=}\Varid{idTree'}\;\Varid{vs}{}\<[E]%
\\
\>[13]{}\mathbf{in}\;\Varid{r}{}\<[E]%
\\[\blanklineskip]%
\>[B]{}\Varid{idTree'}\mathbin{::}\Conid{Tree}\to {\bar{\exists}}\;\Varid{vs}\;.\;(\Varid{vs}\to (\Varid{vs},\Conid{Tree})){}\<[E]%
\\
\>[B]{}\Varid{idTree'}\;(\Conid{Leaf}\;\Varid{v}){}\<[E]%
\\
\>[B]{}\hsindent{5}{}\<[5]%
\>[5]{}\mathrel{=}\lessdot\Conid{Int}, (\lambda \Varid{w}\to {}\<[25]%
\>[25]{}(\Varid{v},\Conid{Leaf}\;\Varid{w}))\gtrdot{}\<[E]%
\\
\>[B]{}\Varid{idTree'}\;(\Conid{Bin}\;\Varid{l}\;\Varid{r}){}\<[E]%
\\
\>[B]{}\hsindent{5}{}\<[5]%
\>[5]{}\mathrel{=}{}\<[8]%
\>[8]{}\lambda {}\<[8E]%
\>[11]{}\mathord{\sim}(\Varid{vsl'},\Varid{vsr'}){}\<[27]%
\>[27]{}\to {}\<[27E]%
\\
\>[11]{}\mathbf{let}\;{}\<[16]%
\>[16]{}\lessdot\Varid{t}_{\Varid{vsl}}, (\Varid{vsl},\Varid{tl})\gtrdot{}\<[39]%
\>[39]{}\mathrel{=}\Varid{idTree'}\;\Varid{l}\;\Varid{vsl'}{}\<[E]%
\\
\>[16]{}\lessdot\Varid{t}_{\Varid{vsr}}, (\Varid{vsr},\Varid{tr})\gtrdot{}\<[39]%
\>[39]{}\mathrel{=}\Varid{idTree'}\;\Varid{r}\;\Varid{vsr'}{}\<[E]%
\\
\>[11]{}\mathbf{in}\;{}\<[16]%
\>[16]{}\lessdot(\Varid{t}_{\Varid{vsl}},\Varid{t}_{\Varid{vsr}}), ((\Varid{vsl},\Varid{vsr}),\Varid{tl}\mathbin{`\Conid{Bin}`}\Varid{tr})\gtrdot{}\<[E]%
\ColumnHook
\end{hscode}\resethooks
\caption{\ensuremath{\Varid{idTree}} using \ensuremath{{\bar{\exists}}}}\label{fig:uhc}
\end{figure}

Before paying attention to the precise type rules relating to
\ensuremath{{\bar{\exists}}} we will give yet another example of its usefulness.

\subsection{\ensuremath{\Varid{sortTree}}}\label{classes}

As a next step in showing that it may be undesirable to separate the
computation of the type from the computation of part of the result we
modify our \ensuremath{\Varid{idTree}} example by requiring that the leaf values from the
original tree are to be reordered in such a way that a prefix
traversal of the resulting tree finds and increasing list of leaf
values; the resulting tree however should have the same shape again.

In order to explain our algorithm we first give a version
(Figure~\ref{fig:listsorting}) in which we use conventional lists to represent
collected leaf values. In order to avoid expensive concatenations we
thread two list values through the tree: one in a backwards direction
in which we collect the leaf values, and one in a forwards direction
from which we take leaf values. At the top level the constructed
first value is used to initialise the second one. 

The helper function \ensuremath{\Varid{sortTree'}} takes as
arguments: 
\begin{itemize}
\item the \ensuremath{\Conid{Tree}} to be sorted
\item the \emph{sorted list} \ensuremath{\Varid{rest}} containing the leaf values
  following the node in a prefix traversal 
\item the list \ensuremath{\Varid{xs}} of values still not used in building the result
  tree. 
\end{itemize}

It returns:
\begin{itemize}
\item a sorted list of leaf values containing the leaf values of
  this node  and the leave values  following the node at hand in a
prefix traversal (i.e. it adds its contained leaves to its second
argument), 
\item a tail of the parameter \ensuremath{\Varid{xs}} containing the values to be used in
  constructing the rest of the tree 
\item  the final tree constructed from the prefix of its \ensuremath{\Varid{xs}} argument
\end{itemize}

\begin{figure}
\begin{hscode}\SaveRestoreHook
\column{B}{@{}>{\hspre}l<{\hspost}@{}}%
\column{3}{@{}>{\hspre}c<{\hspost}@{}}%
\column{3E}{@{}l@{}}%
\column{6}{@{}>{\hspre}l<{\hspost}@{}}%
\column{11}{@{}>{\hspre}l<{\hspost}@{}}%
\column{15}{@{}>{\hspre}l<{\hspost}@{}}%
\column{17}{@{}>{\hspre}l<{\hspost}@{}}%
\column{20}{@{}>{\hspre}l<{\hspost}@{}}%
\column{23}{@{}>{\hspre}l<{\hspost}@{}}%
\column{29}{@{}>{\hspre}l<{\hspost}@{}}%
\column{30}{@{}>{\hspre}l<{\hspost}@{}}%
\column{31}{@{}>{\hspre}l<{\hspost}@{}}%
\column{37}{@{}>{\hspre}l<{\hspost}@{}}%
\column{41}{@{}>{\hspre}l<{\hspost}@{}}%
\column{44}{@{}>{\hspre}l<{\hspost}@{}}%
\column{50}{@{}>{\hspre}l<{\hspost}@{}}%
\column{60}{@{}>{\hspre}l<{\hspost}@{}}%
\column{66}{@{}>{\hspre}l<{\hspost}@{}}%
\column{E}{@{}>{\hspre}l<{\hspost}@{}}%
\>[B]{}\Varid{sortTree}\;\Varid{t}\mathrel{=}{}\<[15]%
\>[15]{}\mathbf{let}\;{}\<[20]%
\>[20]{}(\Varid{vs},[\mskip1.5mu \mskip1.5mu],\Varid{res})\mathrel{=}\Varid{sortTree'}\;\Varid{t}\;[\mskip1.5mu \mskip1.5mu]\;\Varid{vs}{}\<[E]%
\\
\>[15]{}\mathbf{in}\;{}\<[20]%
\>[20]{}\Varid{res}{}\<[E]%
\\[\blanklineskip]%
\>[B]{}\Varid{insert}\;\Varid{v}\;[\mskip1.5mu \mskip1.5mu]\mathrel{=}[\mskip1.5mu \Varid{v}\mskip1.5mu]{}\<[E]%
\\
\>[B]{}\Varid{insert}\;\Varid{v}\;(\Varid{w}\mathbin{:}\Varid{ws})\mathrel{=}\mathbf{if}\;\Varid{v}\mathbin{<}\Varid{w}\;{}\<[31]%
\>[31]{}\mathbf{then}\;{}\<[37]%
\>[37]{}\Varid{v}\mathbin{:}\Varid{w}\mathbin{:}\Varid{ws}{}\<[E]%
\\
\>[31]{}\mathbf{else}\;{}\<[37]%
\>[37]{}\Varid{w}\mathbin{:}\Varid{insert}\;\Varid{v}\;\Varid{ws}{}\<[E]%
\\[\blanklineskip]%
\>[B]{}\Varid{sortTree'}\mathbin{::}\Conid{Tree}\to [\mskip1.5mu \Conid{Int}\mskip1.5mu]\to [\mskip1.5mu \Conid{Int}\mskip1.5mu]\to ([\mskip1.5mu \Conid{Int}\mskip1.5mu],[\mskip1.5mu \Conid{Int}\mskip1.5mu],\Conid{Tree}){}\<[E]%
\\
\>[B]{}\Varid{sortTree'}\;(\Conid{Leaf}\;\Varid{v})\;{}\<[23]%
\>[23]{}\Varid{rest}{}\<[30]%
\>[30]{}\mathord{\sim}(\Varid{x}\mathbin{:}\Varid{xs}){}\<[41]%
\>[41]{}\mathrel{=}(\Varid{insert}\;\Varid{v}\;\Varid{rest},{}\<[60]%
\>[60]{}\Varid{xs},{}\<[66]%
\>[66]{}\Conid{Leaf}\;\Varid{x}){}\<[E]%
\\
\>[B]{}\Varid{sortTree'}\;(\Conid{Bin}\;\Varid{l}\;\Varid{r})\;{}\<[23]%
\>[23]{}\Varid{rest}\;{}\<[30]%
\>[30]{}\Varid{xs}{}\<[E]%
\\
\>[B]{}\hsindent{3}{}\<[3]%
\>[3]{}\mathrel{=}{}\<[3E]%
\>[6]{}\mathbf{let}\;{}\<[11]%
\>[11]{}(\Varid{vl},{}\<[17]%
\>[17]{}\Varid{xsl},{}\<[23]%
\>[23]{}\Varid{tl}){}\<[29]%
\>[29]{}\mathrel{=}\Varid{sortTree'}\;\Varid{l}\;{}\<[44]%
\>[44]{}\Varid{vr}\;{}\<[50]%
\>[50]{}\Varid{xs}{}\<[E]%
\\
\>[11]{}(\Varid{vr},{}\<[17]%
\>[17]{}\Varid{xsr},{}\<[23]%
\>[23]{}\Varid{tr}){}\<[29]%
\>[29]{}\mathrel{=}\Varid{sortTree'}\;\Varid{r}\;{}\<[44]%
\>[44]{}\Varid{rest}\;{}\<[50]%
\>[50]{}\Varid{xsl}{}\<[E]%
\\
\>[6]{}\mathbf{in}\;{}\<[11]%
\>[11]{}(\Varid{vl},{}\<[17]%
\>[17]{}\Varid{xsr},{}\<[23]%
\>[23]{}\Conid{Bin}\;\Varid{tl}\;\Varid{tr}){}\<[E]%
\ColumnHook
\end{hscode}\resethooks
\caption{Sorting using lists}\label{fig:listsorting}
\end{figure}

Our next step (Figure~\ref{fig:sorting}) is to replace the
intermediate lists with nested Cartesian products.  This guarantees
that the top-level \ensuremath{\Varid{sortTree}} function cannot cheat, e.g. by secretly
replacing elements in the list of leaf values, and each node adds
exactly one element to the list and removes exactly one element

The data type \ensuremath{\Conid{Ordlist}} represents a sorted list to which elements can
be added using the function part it carries. This makes more
explicit what is going on. We have used scoped type variables to get
hold of the type of the parameter \ensuremath{\Varid{rest}} and used explicit type
annotations in \ensuremath{\Varid{vr}\mathbin{::}\Varid{t}_{\Varid{r}}} to indicate how the
polymorphic type \ensuremath{\Varid{rest}} in the calls is to be instantiated. Note that
the returned type does not simply depend on the shape of the argument
tree anymore, but also on the polymorphic type \ensuremath{\Varid{rest}}!

\begin{figure}
\begin{hscode}\SaveRestoreHook
\column{B}{@{}>{\hspre}l<{\hspost}@{}}%
\column{3}{@{}>{\hspre}l<{\hspost}@{}}%
\column{6}{@{}>{\hspre}l<{\hspost}@{}}%
\column{7}{@{}>{\hspre}c<{\hspost}@{}}%
\column{7E}{@{}l@{}}%
\column{11}{@{}>{\hspre}l<{\hspost}@{}}%
\column{13}{@{}>{\hspre}l<{\hspost}@{}}%
\column{14}{@{}>{\hspre}l<{\hspost}@{}}%
\column{16}{@{}>{\hspre}l<{\hspost}@{}}%
\column{19}{@{}>{\hspre}l<{\hspost}@{}}%
\column{21}{@{}>{\hspre}l<{\hspost}@{}}%
\column{22}{@{}>{\hspre}l<{\hspost}@{}}%
\column{23}{@{}>{\hspre}l<{\hspost}@{}}%
\column{27}{@{}>{\hspre}l<{\hspost}@{}}%
\column{28}{@{}>{\hspre}l<{\hspost}@{}}%
\column{29}{@{}>{\hspre}l<{\hspost}@{}}%
\column{33}{@{}>{\hspre}l<{\hspost}@{}}%
\column{35}{@{}>{\hspre}l<{\hspost}@{}}%
\column{44}{@{}>{\hspre}l<{\hspost}@{}}%
\column{45}{@{}>{\hspre}l<{\hspost}@{}}%
\column{47}{@{}>{\hspre}l<{\hspost}@{}}%
\column{57}{@{}>{\hspre}l<{\hspost}@{}}%
\column{E}{@{}>{\hspre}l<{\hspost}@{}}%
\>[B]{}\mathbf{data}\;\Conid{OrdList}\;\Varid{cl}\;\mathbf{where}{}\<[E]%
\\
\>[B]{}\hsindent{3}{}\<[3]%
\>[3]{}\Conid{OrdList}\mathbin{::}\Varid{cl}\to (\Conid{Int}\to \Varid{cl}\to (\Conid{Int},\Varid{cl}))\to \Conid{Ordlist}{}\<[E]%
\\[\blanklineskip]%
\>[B]{}\Varid{sortTree}\;\Varid{t}\mathrel{=}{}\<[16]%
\>[16]{}\mathbf{let}\;\lessdot\Varid{t}_{\Varid{vs}}, (\Conid{OrdList}\;\Varid{vs},\anonymous ,\Varid{res})\gtrdot{}\<[E]%
\\
\>[16]{}\hsindent{7}{}\<[23]%
\>[23]{}\mathrel{=}\Varid{sortTree''}\;\Varid{t}\;{}\<[E]%
\\
\>[23]{}\hsindent{6}{}\<[29]%
\>[29]{}(\Conid{OrdList}\;()\;(\lambda \Varid{x}\;()\to (\Varid{x},())))\;\Varid{vs}{}\<[E]%
\\
\>[16]{}\mathbf{in}\;{}\<[21]%
\>[21]{}\Varid{res}{}\<[E]%
\\
\>[B]{}\Varid{sortTree''}{}\<[13]%
\>[13]{}\mathbin{::}\Conid{Tree}{}\<[22]%
\>[22]{}\to \forall\;\Varid{rest}\;.\;\Conid{OrdList}\;\Varid{rest}{}\<[E]%
\\
\>[22]{}\to {\bar{\exists}}\;\Varid{xs}\;.\;\Varid{xs}\to (\Conid{OrdList}\;\Varid{xs},\Varid{rest},\Conid{Tree}){}\<[E]%
\\[\blanklineskip]%
\>[B]{}\Varid{sortTree''}\;(\Conid{Leaf}\;\Varid{v})\;(\Conid{OrdList}\;(\Varid{rest}\mathbin{::}\Varid{t}_{\Varid{rest}})\;\Varid{insert}){}\<[57]%
\>[57]{}\mathord{\sim}(\Varid{x},\Varid{xs}){}\<[E]%
\\
\>[B]{}\hsindent{7}{}\<[7]%
\>[7]{}\mathrel{=}{}\<[7E]%
\>[11]{}\lessdot(\Conid{Int},\Varid{t}_{\Varid{rest}}), {}\<[E]%
\\
\>[11]{}\hsindent{5}{}\<[16]%
\>[16]{}(\Conid{OrdList}\;{}\<[E]%
\\
\>[16]{}\hsindent{3}{}\<[19]%
\>[19]{}(\Varid{ins}\;\Varid{v}\;\Varid{rest})\;{}\<[E]%
\\
\>[16]{}\hsindent{3}{}\<[19]%
\>[19]{}(\lambda \Varid{w}\;(\Varid{x},\Varid{xs})\to {}\<[35]%
\>[35]{}\mathbf{if}\;\Varid{w}\mathbin{<}\Varid{x}\;\mathbf{then}\;(\Varid{w},(\Varid{x},\Varid{xs})){}\<[E]%
\\
\>[35]{}\hsindent{10}{}\<[45]%
\>[45]{}\mathbf{else}\;(\Varid{x},\Varid{insert}\;\Varid{w}\;\Varid{xs})){}\<[E]%
\\
\>[11]{},\Varid{xs},\Conid{Leaf}\;\Varid{x})\gtrdot{}\<[E]%
\\[\blanklineskip]%
\>[B]{}\Varid{sortTree''}\;(\Conid{Bin}\;\Varid{l}\;\Varid{r})\;(\Varid{rest}\mathbin{::}\Varid{t}_{\Varid{rest}})\;\Varid{xs}{}\<[E]%
\\
\>[B]{}\hsindent{3}{}\<[3]%
\>[3]{}\mathrel{=}{}\<[6]%
\>[6]{}\mathbf{let}\;{}\<[11]%
\>[11]{}\lessdot\Varid{t}_{\Varid{l}}, {}\<[21]%
\>[21]{}(\Varid{vl},{}\<[27]%
\>[27]{}\Varid{xsl},{}\<[33]%
\>[33]{}\Varid{tl})\gtrdot{}\<[E]%
\\
\>[11]{}\hsindent{3}{}\<[14]%
\>[14]{}\mathrel{=}\Varid{sortTree''}\;{}\<[28]%
\>[28]{}\Varid{l}\;(\Varid{vr}\mathbin{::}\Varid{t}_{\Varid{r}})\;{}\<[44]%
\>[44]{}\Varid{xs}{}\<[E]%
\\
\>[11]{}\lessdot\Varid{t}_{\Varid{r}}, {}\<[21]%
\>[21]{}(\Varid{vr},{}\<[27]%
\>[27]{}\Varid{xsr},{}\<[33]%
\>[33]{}\Varid{tr})\gtrdot{}\<[E]%
\\
\>[11]{}\hsindent{3}{}\<[14]%
\>[14]{}\mathrel{=}\Varid{sortTree''}\;{}\<[28]%
\>[28]{}\Varid{r}\;(\Varid{rest}\mathbin{::}\Varid{t}_{\Varid{rest}})\;{}\<[47]%
\>[47]{}\Varid{xsl}{}\<[E]%
\\
\>[6]{}\mathbf{in}\;{}\<[11]%
\>[11]{}\lessdot\Varid{t}_{\Varid{l}}, {}\<[21]%
\>[21]{}(\Varid{vl},{}\<[27]%
\>[27]{}\Varid{xsr},{}\<[33]%
\>[33]{}\Conid{Bin}\;\Varid{tl}\;\Varid{tr})\gtrdot{}\<[E]%
\ColumnHook
\end{hscode}\resethooks
\caption{Sorting using Cartesian products}\label{fig:sorting}
\end{figure}

\section{Polymorphic Contexts}\label{polycontexts}

\subsection{The \ensuremath{{\bar{\exists}}} quantifier}

By inspecting at the uses of \ensuremath{{\bar{\exists}}} in both examples we see that
the types constructed do not play a role at all at the place where
they are fed back into the computation; they only serve to describe
the type of part of the result of the function which is used to pass
back an argument of that type.  This suggests that it is more
natural to compute this type as part of the result and use a
pattern-matching \ensuremath{\mathbf{letrec}} construct to get access to this type; thus
making it possible to enforce that the value passed back is described
by the computed type. It is here that we deviate from the standard way
of dealing with existential values. We observe that once all arguments to the
function call have been given the computed type in principle is fully
determined, and could be computed by the callee. This might however
imply that we have to split the computation: one part in which we just
do enough work to compute the type, and once the type has been determined
the rest of the computation to compute its associated value, as in
Figure \ref{fig:idTreeppghc}. 
We thus distinguish between a type being
\emph{computable}, meaning that all information needed to compute it is
available, and \emph{computed}, meaning that it has been
computed from this available information.

This brings us to the most important message of this paper;  
we are dealing with a situation in which 
\emph{the r\^oles of the context in which a function is called 
and the function itself are reversed with respect to which
side of the function call decides how types are to be instantiated}:
\begin{itemize}
\item when calling a  \emph{polymorphic function} it is the \emph{context} which
  decides on the type of the polymorphic argument and it is the
  duty of the function to return a value consistent with that type.
\item a \emph{function} requiring a \emph{polymorphic context} decides
  on the type returned by the call and it is the duty of the context
  to pass back an argument consistent with that type.
\end{itemize}

For the callee it looks like calling into a polymorphic
context. In our \ensuremath{\Varid{idTree}} example the context behaves as
a function of type \ensuremath{\forall\;\Varid{a}\;.\;\Varid{a}\to \Varid{a}}. 

We may wonder what is the correct place to insert the \ensuremath{{\bar{\exists}}}
quantifier. Just as the type rules of System-F can be used to show
that there is no essential difference between the types \ensuremath{\forall\;\Varid{a}\;.\;\Conid{Int}\to \Varid{a}\to \Varid{a}} and \ensuremath{\Conid{Int}\to \forall\;\Varid{a}\;.\;\Varid{a}\to \Varid{a}} our rules will make that we do not
have to distinguish between \ensuremath{{\bar{\exists}}\;\Varid{t}_{\Varid{vs}}\;.\;\Conid{Tree}\to \Varid{t}_{\Varid{vs}}\to (\Varid{t}_{\Varid{vs}},\Conid{Tree})} and \ensuremath{\Conid{Tree}\to {\bar{\exists}}\;\Varid{t}_{\Varid{vs}}\;.\;\Varid{t}_{\Varid{vs}}\to (\Varid{t}_{\Varid{vs}},\Varid{tree})}.

\subsection{Type rules}\label{typerules}

\begin{TabularCenterFigure}{}{Structures}{syntax}{p{0.99\linewidth}}
  \parbox{0.99\linewidth}{%
\begin{hscode}\SaveRestoreHook
\column{B}{@{}>{\hspre}l<{\hspost}@{}}%
\column{13}{@{}>{\hspre}c<{\hspost}@{}}%
\column{13E}{@{}l@{}}%
\column{20}{@{}>{\hspre}l<{\hspost}@{}}%
\column{26}{@{}>{\hspre}l<{\hspost}@{}}%
\column{29}{@{}>{\hspre}l<{\hspost}@{}}%
\column{53}{@{}>{\hspre}l<{\hspost}@{}}%
\column{E}{@{}>{\hspre}l<{\hspost}@{}}%
\>[B]{}e{}\<[13]%
\>[13]{}\mathrel{=}{}\<[13E]%
\>[20]{}v{}\<[53]%
\>[53]{}\; \mbox{\onelinecomment  variable}{}\<[E]%
\\
\>[13]{}\mid {}\<[13E]%
\>[20]{}e\; e{}\<[53]%
\>[53]{}\; \mbox{\onelinecomment  application}{}\<[E]%
\\
\>[13]{}\mid {}\<[13E]%
\>[20]{}\lambda (v\mathbin{:}\sigma)\to e{}\<[53]%
\>[53]{}\; \mbox{\onelinecomment  abstraction}{}\<[E]%
\\
\>[13]{}\mid {}\<[13E]%
\>[20]{}{\bar{\exists}}\;\alpha\;.\;e{}\<[53]%
\>[53]{}\; \mbox{\onelinecomment  introduction}{}\<[E]%
\\
\>[13]{}\mid {}\<[13E]%
\>[20]{}e\;[\mskip1.5mu \nu\mskip1.5mu]{}\<[53]%
\>[53]{}\; \mbox{\onelinecomment  elimination}{}\<[E]%
\\
\>[13]{}\mid {}\<[13E]%
\>[20]{}\lessdot(\alpha\mathrel{=}\sigma), \Varid{e}\gtrdot{}\<[53]%
\>[53]{}\; \mbox{\onelinecomment  pack}{}\<[E]%
\\
\>[13]{}\mid {}\<[13E]%
\>[20]{}e_{\mathrm{1}}\;\talloblong\mathbin{...}\talloblong\;e_{\Varid{n}}{}\<[53]%
\>[53]{}\; \mbox{\onelinecomment  function alternatives}{}\<[E]%
\\
\>[13]{}\mid {}\<[13E]%
\>[20]{}\mathbin{...}{}\<[53]%
\>[53]{}\; \mbox{\onelinecomment  other terms}{}\<[E]%
\\
\>[B]{}\sigma{}\<[13]%
\>[13]{}\mathrel{=}{}\<[13E]%
\>[20]{}\alpha{}\<[53]%
\>[53]{}\; \mbox{\onelinecomment  type variable}{}\<[E]%
\\
\>[13]{}\mid {}\<[13E]%
\>[20]{}\nu{}\<[53]%
\>[53]{}\; \mbox{\onelinecomment  type name from the program}{}\<[E]%
\\
\>[13]{}\mid {}\<[13E]%
\>[20]{}\sigma\to \sigma{}\<[53]%
\>[53]{}\; \mbox{\onelinecomment  abstraction}{}\<[E]%
\\
\>[13]{}\mid {}\<[13E]%
\>[20]{}\forall\;{}\<[29]%
\>[29]{}\alpha\;.\;\sigma{}\<[53]%
\>[53]{}\; \mbox{\onelinecomment  universal quantification}{}\<[E]%
\\
\>[13]{}\mid {}\<[13E]%
\>[20]{}{\bar{\exists}}\;{}\<[29]%
\>[29]{}\alpha\;.\;\sigma{}\<[53]%
\>[53]{}\; \mbox{\onelinecomment  our existential}{}\<[E]%
\\
\>[13]{}\mid {}\<[13E]%
\>[20]{}\lessdot{}\<[26]%
\>[26]{}(\alpha\mathrel{=}\sigma), \sigma\gtrdot{}\<[53]%
\>[53]{}\; \mbox{\onelinecomment  packed existential}{}\<[E]%
\\
\>[13]{}\mid {}\<[13E]%
\>[20]{}\mathbin{...}{}\<[53]%
\>[53]{}\; \mbox{\onelinecomment  other types}{}\<[E]%
\ColumnHook
\end{hscode}\resethooks
}
\end{TabularCenterFigure}

In Figure ~\ref{syntax} we have given the underlying syntax of our
extension, as far as it deviates from standard System-F. We have made the underlying typing a lot more explicit than 
in the example code thus far, which was made to resemble Haskell as much as
possible. In practice a lot of this
information can be inferred, as we are assuming in a lot of our
example code. 

In Figure \ref{fig:explicit} we have rephrased our \ensuremath{\Varid{idTree}}
function once more, but now with more explicit typing directives, and
making clear how function alternatives can be represented. Our
function \ensuremath{\Varid{idTree'}} is composed of alternatives \ensuremath{\Varid{idTree'}_{\Conid{Leaf}}} and
\ensuremath{\Varid{idTree'}_{\Conid{Bin}}}, each defined as a
\ensuremath{{\bar{\exists}}}-quantified \ensuremath{\lambda }-expression. We require that all parameters
with a type introduced by the \ensuremath{{\bar{\exists}}} match lazily, so they cannot
be inspected at runtime when matching the pattern. The
alternatives are tried in order until a match occurs. The result of
the function is constructed by explicitly packing a value with the
corresponding type. 

If we look at the definition of \ensuremath{\Varid{idTree}} we see that the expression
returns an existentially typed value; the type part is given the name
\ensuremath{\Varid{t}_{\Varid{vs}}}, and this is also the type which is to be used in typing the
expressions. Since this very much resembles the way polymorphic
functions are instantiated we have again borrowed notation from
System-F in \ensuremath{\Varid{idTree'}\;[\mskip1.5mu \Varid{t}_{\Varid{vs}}\mskip1.5mu]}. To paraphrase Henry Ford\footnote{``A
  customer can have a car painted 
any color he wants as long as it is black''} we say that the calling environment can pick any type as long
as it is the returned  \ensuremath{\Varid{t}_{\Varid{vs}}}, because that is the only way in which the returned result can
match the left-hand side \ensuremath{\lessdot\Varid{t}_{\Varid{vs}}, (\Varid{vs},\Varid{r})\gtrdot}.  

\begin{figure}
\begin{hscode}\SaveRestoreHook
\column{B}{@{}>{\hspre}l<{\hspost}@{}}%
\column{13}{@{}>{\hspre}l<{\hspost}@{}}%
\column{15}{@{}>{\hspre}c<{\hspost}@{}}%
\column{15E}{@{}l@{}}%
\column{18}{@{}>{\hspre}l<{\hspost}@{}}%
\column{21}{@{}>{\hspre}l<{\hspost}@{}}%
\column{26}{@{}>{\hspre}l<{\hspost}@{}}%
\column{31}{@{}>{\hspre}l<{\hspost}@{}}%
\column{32}{@{}>{\hspre}l<{\hspost}@{}}%
\column{35}{@{}>{\hspre}l<{\hspost}@{}}%
\column{39}{@{}>{\hspre}l<{\hspost}@{}}%
\column{46}{@{}>{\hspre}l<{\hspost}@{}}%
\column{49}{@{}>{\hspre}l<{\hspost}@{}}%
\column{52}{@{}>{\hspre}c<{\hspost}@{}}%
\column{52E}{@{}l@{}}%
\column{55}{@{}>{\hspre}l<{\hspost}@{}}%
\column{62}{@{}>{\hspre}l<{\hspost}@{}}%
\column{74}{@{}>{\hspre}l<{\hspost}@{}}%
\column{E}{@{}>{\hspre}l<{\hspost}@{}}%
\>[B]{}\Varid{idTree}\;\Varid{t}\mathrel{=}{}\<[13]%
\>[13]{}\mathbf{let}\;\lessdot\Varid{t}_{\Varid{vs}}, (\Varid{vs},\Varid{r})\gtrdot\mathrel{=}\Varid{idTree'}\;\Varid{vs}{}\<[E]%
\\
\>[13]{}\mathbf{in}\;\Varid{r}{}\<[E]%
\\[\blanklineskip]%
\>[B]{}\Varid{idTree}\mathbin{::}\Conid{Tree}\to \Conid{Tree}{}\<[E]%
\\
\>[B]{}\Varid{idTree}\;\Varid{t}\mathrel{=}{}\<[13]%
\>[13]{}\mathbf{let}\;\lessdot\Varid{t}_{\Varid{vs}}, (\Varid{vs},\Varid{r})\gtrdot\mathrel{=}\Varid{idTree'}\;[\mskip1.5mu \Varid{t}_{\Varid{vs}}\mskip1.5mu]\;\Varid{t}\;\Varid{vs}\;\mathbf{in}\;{}\<[62]%
\>[62]{}\Varid{r}{}\<[E]%
\\[\blanklineskip]%
\>[B]{}\Varid{idTree'}\mathbin{::}{\bar{\exists}}\;\Varid{t}_{\Varid{vs}}\;.\;\Conid{Tree}\to \Varid{t}_{\Varid{vs}}\to (\Varid{t}_{\Varid{vs}},\Conid{Tree}){}\<[E]%
\\
\>[B]{}\Varid{idTree'}\mathrel{=}\Varid{idTree'}_{\Conid{Leaf}}\;\talloblong\;\Varid{idTree'}_{\Conid{Bin}}{}\<[E]%
\\[\blanklineskip]%
\>[B]{}\Varid{idTree'}_{\Conid{Leaf}}{}\<[15]%
\>[15]{}\mathrel{=}{}\<[15E]%
\>[18]{}{\bar{\exists}}\;\Varid{t}_{\Varid{vs}}\;.\;{}\<[E]%
\\
\>[18]{}\lambda {}\<[21]%
\>[21]{}(\Conid{Leaf}\;\Varid{v})\;{}\<[31]%
\>[31]{}\Varid{w}\to \lessdot(\Varid{t}_{\Varid{vs}}\mathrel{=}\Conid{Int}), (\Varid{v},\Conid{Leaf}\;\Varid{w})\gtrdot{}\<[E]%
\\
\>[B]{}\Varid{idTree'}_{\Conid{Bin}}{}\<[15]%
\>[15]{}\mathrel{=}{}\<[15E]%
\>[18]{}{\bar{\exists}}\;\Varid{t}_{\Varid{vs}}\;.\;{}\<[E]%
\\
\>[18]{}\lambda {}\<[21]%
\>[21]{}(\Conid{Bin}\;\Varid{l}\;\Varid{r})\; {}\<[35]%
\>[35]{}\mathord{\sim}(\Varid{vsl'},\Varid{vsr'})\to {}\<[E]%
\\
\>[21]{}\mathbf{let}\;{}\<[26]%
\>[26]{}\lessdot{}\<[32]%
\>[32]{}\Varid{t}_{\Varid{vsl}}, {}\<[39]%
\>[39]{}(\Varid{vsl},\Varid{tl}{}\<[49]%
\>[49]{})\gtrdot{}\<[52]%
\>[52]{}\mathrel{=}{}\<[52E]%
\>[55]{}\Varid{idTree'}\;[\mskip1.5mu \Varid{t}_{\Varid{vsl}}\mskip1.5mu]\;\Varid{l}\;{}\<[74]%
\>[74]{}\Varid{vsl'}{}\<[E]%
\\
\>[26]{}\lessdot{}\<[32]%
\>[32]{}\Varid{t}_{\Varid{vsr}}, {}\<[39]%
\>[39]{}(\Varid{vsr},\Varid{tr}{}\<[49]%
\>[49]{})\gtrdot{}\<[52]%
\>[52]{}\mathrel{=}{}\<[52E]%
\>[55]{}\Varid{idTree'}\;[\mskip1.5mu \Varid{t}_{\Varid{vsr}}\mskip1.5mu]\;\Varid{r}\;{}\<[74]%
\>[74]{}\Varid{vsr'}{}\<[E]%
\\
\>[21]{}\mathbf{in}\;{}\<[26]%
\>[26]{}\lessdot{}\<[32]%
\>[32]{}(\Varid{t}_{\Varid{vs}}\mathrel{=}(\Varid{t}_{\Varid{vsl}},\Varid{t}_{\Varid{vsr}})), {}\<[E]%
\\
\>[32]{}((\Varid{vsl},\Varid{vsr}),{}\<[46]%
\>[46]{}\Varid{tl}\mathbin{`\Conid{Bin}`}\Varid{tr})\gtrdot{}\<[E]%
\ColumnHook
\end{hscode}\resethooks
\caption{annotated \ensuremath{\Varid{idTree}} using \ensuremath{{\bar{\exists}}}}\label{fig:explicit}
\end{figure}
 
\rulerCmdUse{typesystem.B.p}
We start out with a new form of judgement for dealing with patterns in
the environment. We extend the rules given by Harper \cite{Harper:2012:PFP:2431407}
with the form needed for our specific purpose, i.e. dealing with
values packed with an existential type. To introduce notation we show the rule \ruleName{varpat}
which states that a variable is a proper pattern and adds a binding \ensuremath{v\mapsto \sigma} to the environment. 
The rule  \ruleName{unpackpat} states that if a pattern \ensuremath{\Varid{p}}  introduces
names in an environment \ensuremath{\Lambda} then so does \ensuremath{\lessdot\nu, \Varid{p}\gtrdot}. The
existential type is given a name \ensuremath{\nu}, which is stored in the
environment, in which it has to be unique.

\rulerCmdUse{typesystem.B.exp}
In Figure~\ref{typesystem.B.exp} we have given the
rules for our form of existential types. 
We are intentionally incomplete as
we only wish to clarify the non-standard construct \ensuremath{{\bar{\exists}}\;\alpha\;.\;\sigma}
for our existential types in the context of the given examples.  Our
conjecture is that System-F cannot deal with the computational order in which
types are computed (as a result) and passed back (into an earlier
parameter), and hence we do not provide a translation to System-F.
This is due to the System-F approach to unpacking, which uses a
continuation style of formulation in e.g. Harper's book \cite{Harper:2012:PFP:2431407} \ensuremath{{\mathbf{open}}\;\Varid{e}\;{\mathbf{as}}\;\Varid{t}\;{\mathbf{with}}\;\Varid{x}\mathbin{:}\sigma\;\mathbf{in}\;\Varid{e'}},
which does not allow \ensuremath{\Varid{x}} to be referred to in \ensuremath{\Varid{e}}.  

Both the language for expressions \ensuremath{e} and
types \ensuremath{\sigma} are open ended, describing the subset of what Haskell
offers required for our examples.
We furthermore require that all
 environments and the types therein are well-formed.

The rule \ruleName{pack} describes how we can forget (part of) a type
and replace it by an existential type. 

Rule \ruleName{E.I} forgets about type \ensuremath{\sigma'} in
\ensuremath{\sigma} by replacing it by a type variable \ensuremath{\alpha}; rule
\ruleName{E.E} does the inverse, by reintroducing the forgotten
type. Since we have no longer access to the original type we assume
this to be the name of some anonymous type with name \ensuremath{\nu}.

To show how these rules can be used in
typing our desired form of \ensuremath{\Varid{idTree}} we consider the various
function alternatives.
Due to the \ruleName{choice} rule, both alternatives must have type \ensuremath{{\bar{\exists}}\;\Varid{t}_{\Varid{vs}}\;.\;\Conid{Tree}\to \Varid{t}_{\Varid{vs}}\to (\Varid{t}_{\Varid{vs}},\Conid{Tree})}. We sketch the derivation of the first alternative: 

$\inferrule*[Right=\ruleName{E.I}]{\inferrule*[Right=\ruleName{abs(*2)}]{\inferrule*[Right=\ruleName{pack}]{\inferrule*[Right=\ruleName{var(*2)}]{ }{\ensuremath{\Gamma,\Varid{v}\mapsto \Conid{Int},\Varid{w}\mapsto \Conid{Int}\vdash } ^{ \ensuremath{e}}  \ensuremath{(\Varid{v},\Conid{Leaf}\;\Varid{w})\mathbin{:}(\Conid{Int},\Conid{Tree})}}} {\ensuremath{\Gamma,\Varid{v}\mapsto \Conid{Int},\Varid{w}\mapsto \Conid{Int}\vdash } ^{ \ensuremath{e}}  \ensuremath{\lessdot(\Varid{t}_{\Varid{vs}}\mathrel{=}\Conid{Int}), (\Varid{v},\Conid{Leaf}\;\Varid{w})\gtrdot} \\ \ensuremath{\mathbin{:}\lessdot(\Varid{t}_{\Varid{vs}}\mathrel{=}\Conid{Int}), (\Varid{t}_{\Varid{vs}},\Conid{Tree})\gtrdot}}} {\ensuremath{\Gamma\vdash } ^{ \ensuremath{e}}  \ensuremath{\lambda (\Conid{Leaf}\;\Varid{v})\;\Varid{w}\to \lessdot(\Varid{t}_{\Varid{vs}}\mathrel{=}\Conid{Int}), (\Varid{v},\Conid{Leaf}\;\Varid{w})\gtrdot} \\ \ensuremath{\mathbin{:}\Conid{Tree}\to \Conid{Int}\to \lessdot(\Varid{t}_{\Varid{vs}}\mathrel{=}\Conid{Int}), (\Varid{t}_{\Varid{vs}},\Conid{Tree})\gtrdot}}} {\ensuremath{\Gamma\vdash } ^{ \ensuremath{e}}  \ensuremath{{\bar{\exists}}\;\Varid{t}_{\Varid{vs}}\;.\;\lambda (\Conid{Leaf}\;\Varid{v})\;\Varid{w}\to \lessdot(\Varid{t}_{\Varid{vs}}\mathrel{=}\Conid{Int}), (\Varid{v},\Conid{Leaf}\;\Varid{w})\gtrdot} \\ \ensuremath{\mathbin{:}{\bar{\exists}}\;\Varid{t}_{\Varid{vs}}\;.\;\Conid{Tree}\to \Varid{t}_{\Varid{vs}}\to (\Varid{t}_{\Varid{vs}},\Conid{Tree})}}\\$
If we read the derivation bottom-up, by applying the introduction rule
\ruleName{E.I} we derive the type of the lambda abstraction to be
\ensuremath{\Conid{Tree}\to \Conid{Int}\to \lessdot(\Varid{t}_{\Varid{vs}}\mathrel{=}\Conid{Int}), (\Varid{t}_{\Varid{vs}},\Conid{Tree})\gtrdot}.
Then, with the application of the usual rule for abstraction (\ruleName{abs})
twice we determine that the type of \ensuremath{\lessdot(\Varid{t}_{\Varid{vs}}\mathrel{=}\Conid{Int}), (\Varid{v},\Conid{Leaf}\;\Varid{w})\gtrdot} is
\ensuremath{\lessdot(\Varid{t}_{\Varid{vs}}\mathrel{=}\Conid{Int}), (\Varid{t}_{\Varid{vs}},\Conid{Tree})\gtrdot} given that \ensuremath{\Varid{v}} and \ensuremath{\Varid{w}} are bound to
\ensuremath{\Conid{Int}} in the environment.
Applying \ruleName{pack} we conclude that this type is correct
given that the pair \ensuremath{(\Varid{v},\Conid{Leaf}\;\Varid{w})} has type \ensuremath{(\Conid{Int},\Conid{Tree})}.

In the second branch, the type of the lambda abstraction is
\ensuremath{\Conid{Tree}\to (\Varid{t}_{\Varid{vsl}},\Varid{t}_{\Varid{vsr}})\to \lessdot(\Varid{t}_{\Varid{vs}}\mathrel{=}(\Varid{t}_{\Varid{vsl}},\Varid{t}_{\Varid{vsr}})), (\Varid{t}_{\Varid{vs}},\Conid{Tree})\gtrdot},
with \ensuremath{\Varid{t}_{\Varid{vsl}}} and \ensuremath{\Varid{t}_{\Varid{vsr}}} coming from the recursive calls of \ensuremath{\Varid{idTree'}}.  
We show how the rules are applied in the derivations corresponding to
the first binding of the \ensuremath{\mathbf{let}} expression.

The derivation for the pattern:

$\inferrule*[Right=\ruleName{unpackpat}]{\inferrule*[Right=\ruleName{varpat}(*2)]{ } {\ensuremath{\Varid{vsl}\mapsto \Varid{t}_{\Varid{vsl}},\Varid{tl}\mapsto \Conid{Tree},\Varid{t}_{\Varid{vsl}}\vdash } ^{ \ensuremath{\Varid{p}}}  \ensuremath{(\Varid{vsl},\Varid{tl})\mathbin{:}(\Varid{t}_{\Varid{vsl}},\Conid{Tree})}}} {\ensuremath{\Varid{vsl}\mapsto \Varid{t}_{\Varid{vsl}},\Varid{tl}\mapsto \Conid{Tree},\Varid{t}_{\Varid{vsl}}\vdash } ^{ \ensuremath{\Varid{p}}}  \ensuremath{\lessdot\Varid{t}_{\Varid{vsl}}, (\Varid{vsl},\Varid{tl})\gtrdot\mathbin{:}} \\ \ensuremath{\lessdot(\Varid{t}_{\Varid{vs}}\mathrel{=}\Varid{t}_{\Varid{vsl}}), (\Varid{t}_{\Varid{vs}},\Conid{Tree})\gtrdot}}\\$
If we read the pattern judgment \ensuremath{\ensuremath{\Lambda\vdash } ^{ \ensuremath{\Varid{p}}} \ensuremath{\Varid{p}\mathbin{:}\sigma}} as
an output \ensuremath{\Lambda} produced out of the inputs \ensuremath{\Varid{p}}  and \ensuremath{\sigma},
we can see how the existential type is unpacked by giving the name \ensuremath{\Varid{t}_{\Varid{vsl}}},
and the pair pattern introduces the bindings \ensuremath{\Varid{vsl}\mapsto \Varid{t}_{\Varid{vsl}}}
and \ensuremath{\Varid{tl}\mapsto \Conid{Tree}}.

The derivation for the right hand side, assume \ensuremath{\Gamma} includes \ensuremath{\Varid{idTree'}\mapsto {\bar{\exists}}\;\Varid{t}_{\Varid{vs}}\;.\;\Conid{Tree}\to \Varid{t}_{\Varid{vs}}\to (\Varid{t}_{\Varid{vs}},\Conid{Tree}),\Varid{l}\mapsto \Conid{Tree},\Varid{vsl'}\mapsto \Varid{t}_{\Varid{vsl}},\Varid{vst}\mapsto \Varid{t}_{\Varid{vsl}},\Varid{lt}\mapsto \Conid{Tree},\Varid{t}_{\Varid{vsl}}}:

$\inferrule*[Right=\ruleName{app-var}]{\inferrule*[Right=\ruleName{app-var}]{\inferrule*[Right=\ruleName{E.E}]{\inferrule*[Right=\ruleName{var}]{ }{\ensuremath{\Gamma\vdash } ^{ \ensuremath{e}}  \ensuremath{\Varid{idTree'}\mathbin{:}{\bar{\exists}}\;\Varid{t}_{\Varid{vs}}\;.\;\Conid{Tree}\to \Varid{t}_{\Varid{vs}}\to (\Varid{t}_{\Varid{vs}},\Conid{Tree})} }} {\ensuremath{\Gamma\vdash } ^{ \ensuremath{e}}  \ensuremath{\Varid{idTree'}\;[\mskip1.5mu \Varid{t}_{\Varid{vsl}}\mskip1.5mu]\mathbin{:}} \\ \ensuremath{\Conid{Tree}\to \Varid{t}_{\Varid{vsl}}\to \lessdot(\Varid{t}_{\Varid{vs}}\mathrel{=}\Varid{t}_{\Varid{vsl}}), (\Varid{t}_{\Varid{vs}},\Conid{Tree})\gtrdot}} } {\ensuremath{\Gamma\vdash } ^{ \ensuremath{e}}  \ensuremath{\Varid{idTree'}\;[\mskip1.5mu \Varid{t}_{\Varid{vsl}}\mskip1.5mu]\;\Varid{l}\mathbin{:}\Varid{t}_{\Varid{vsl}}\to \lessdot(\Varid{t}_{\Varid{vs}}\mathrel{=}\Varid{t}_{\Varid{vsl}}), (\Varid{t}_{\Varid{vs}},\Conid{Tree})\gtrdot}}} {\ensuremath{\Gamma\vdash } ^{ \ensuremath{e}}  \ensuremath{\Varid{idTree'}\;[\mskip1.5mu \Varid{t}_{\Varid{vsl}}\mskip1.5mu]\;\Varid{l}\;\Varid{vsl'}\mathbin{:}\lessdot(\Varid{t}_{\Varid{vs}}\mathrel{=}\Varid{t}_{\Varid{vsl}}), (\Varid{t}_{\Varid{vs}},\Conid{Tree})\gtrdot}}\\$
In this case we apply the usual rules for \ruleName{app} and \ruleName{var} (which we applied together under the name \ruleName{app-var}) twice.
Then, by aplying the elimination \ruleName{E.E},
we conclude that \ensuremath{\Varid{idTree'}} has type
\ensuremath{{\bar{\exists}}\;\Varid{t}_{\Varid{vs}}\;.\;\Conid{Tree}\to \Varid{t}_{\Varid{vs}}\to (\Varid{t}_{\Varid{vs}},\Conid{Tree})}, which can be unified to the type
we already have in the context.

\subsection{Safety}
We want to stress the differences between our \ensuremath{{\bar{\exists}}} and the
conventional \ensuremath{\exists}. This is shown by the following program:

\begin{hscode}\SaveRestoreHook
\column{B}{@{}>{\hspre}l<{\hspost}@{}}%
\column{4}{@{}>{\hspre}c<{\hspost}@{}}%
\column{4E}{@{}l@{}}%
\column{6}{@{}>{\hspre}l<{\hspost}@{}}%
\column{12}{@{}>{\hspre}l<{\hspost}@{}}%
\column{15}{@{}>{\hspre}l<{\hspost}@{}}%
\column{17}{@{}>{\hspre}l<{\hspost}@{}}%
\column{18}{@{}>{\hspre}l<{\hspost}@{}}%
\column{22}{@{}>{\hspre}l<{\hspost}@{}}%
\column{24}{@{}>{\hspre}l<{\hspost}@{}}%
\column{E}{@{}>{\hspre}l<{\hspost}@{}}%
\>[B]{}\Varid{f}\mathbin{::}\exists\;\Varid{x}\;.\;\Varid{x}\to \Conid{Bool}\to (\Varid{x},\Conid{Int}){}\<[E]%
\\
\>[B]{}\Varid{f}{}\<[4]%
\>[4]{}\mathrel{=}{}\<[4E]%
\>[12]{}\lambda {}\<[15]%
\>[15]{}\Varid{v}\;{}\<[18]%
\>[18]{}\Conid{True}{}\<[24]%
\>[24]{}\to (\Varid{v}\mathbin{+}\mathrm{1},\Varid{v}){}\<[E]%
\\
\>[4]{}\talloblong{}\<[4E]%
\>[12]{}\lambda \Varid{c}\;{}\<[17]%
\>[17]{}\Conid{False}{}\<[24]%
\>[24]{}\to (\Varid{chr}\;(\Varid{ord}\;\Varid{c}\mathbin{+}\mathrm{1}),\Varid{ord}\;\Varid{c}){}\<[E]%
\\[\blanklineskip]%
\>[B]{}\mathbf{let}\;{}\<[6]%
\>[6]{}\lessdot\Varid{t}_{\Varid{c}}, \Varid{g}\gtrdot\mathrel{=}\Varid{f}{}\<[22]%
\>[22]{}\mbox{\onelinecomment  unpacking f }{}\<[E]%
\\
\>[6]{}(\Varid{i},\Varid{v1})\mathrel{=}\Varid{g}\;\Varid{c}\;\Conid{True}{}\<[E]%
\\
\>[6]{}(\Varid{c},\Varid{v2})\mathrel{=}\Varid{g}\;\Varid{i}\;\Conid{False}{}\<[E]%
\\
\>[B]{}\mathbf{in}\mathbin{...}{}\<[E]%
\ColumnHook
\end{hscode}\resethooks

Here we instantiate the existential type of \ensuremath{\Varid{f}} with a
type constant and then bind the resulting value to a \ensuremath{\Varid{g}}.  Both calls
of \ensuremath{\Varid{g}} are now assuming that the same type constant is used for the
existential value in the type of \ensuremath{\Varid{f}}, causing type checking to succeed
where its should not. With our \ensuremath{{\bar{\exists}}} this cannot happen since we
cannot unpack \ensuremath{\Varid{f}} but only a result. Suppose \ensuremath{\Varid{f}} has type \ensuremath{{\bar{\exists}}\;\Varid{x}\;.\;\Varid{x}\to \Conid{Bool}\to (\Varid{x},\Conid{Int})},
if we want to unpack it we have to start by eliminating the \ensuremath{{\bar{\exists}}}
by doing \ensuremath{\Varid{f}\;[\mskip1.5mu \Varid{t}_{\Varid{c}}\mskip1.5mu]}.
This expression has type \ensuremath{\Varid{t}_{\Varid{c}}\to \Conid{Bool}\to \lessdot(\alpha\mathrel{=}\Varid{t}_{\Varid{c}}), (\alpha,\Conid{Int})\gtrdot}
and thus it cannot be used as right hand side of a let
binding. Apparently the \ruleName{Choice} cannot be applied to combine
values of conventional existential types. So we have decided to leave
them out since they can be emulated using \ensuremath{\forall}.


\section{The \ensuremath{\Conid{ST}} monad}\label{stmonad}
In our last example we will show how we can put the fact that we
have made it possible to have lazy unpacking to further good use. We
will use the type stemming from the unpacking match to parameterise a
polymorphic function, which returns that existentially
typed value!

The state monad \ensuremath{\Conid{ST}} \cite{Launchbury:1994:LFS:178243.178246} is an
important Haskell data type, and a de-facto required part of any
Haskell infrastructure.  Although there are good reasons for
supporting this data type at a very low level, and for providing it
with extensive runtime support, the question arises whether we can
implement the data type in Haskell itself.  


The \ensuremath{\Conid{ST}}-monad represents a \emph{stateful computation}; i.e. a
computation that takes a state and transforms it into another state.
Such transformations include extending the state by introducing a new
variable, and writing and reading already introduced variables.  The
code in Figure \ref{fig:stcode}   introduces the types of the corresponding
basic operations \ensuremath{\Varid{newSTRef}}, \ensuremath{\Varid{writeSTRef}} and \ensuremath{\Varid{readSTRef}}.  

The function \ensuremath{\Varid{runST}} creates a new empty state, runs the
computation starting with this new state, discards the final state
when done and \ensuremath{\Varid{return}}s the value resulting from the computation.
Its type \ensuremath{(\forall\;\Varid{a}\;.\;\Conid{ST}\;\Varid{s}\;\Varid{a})\to \Varid{a}} is interesting by itself because of
its higher order type: it takes a monad \ensuremath{\Conid{ST}\;\Varid{s}\;\Varid{a}} that is polymorphic
in \ensuremath{\Varid{s}}. One might think of this as giving the function \ensuremath{\Varid{runST}} the right
to choose a unique label for a new `named heap' by instantiating the
type \ensuremath{\Varid{s}}, the choice of which is kept hidden from the rest of the program.
This label is then used to label the handed out references of type \ensuremath{\Conid{STRef}\;\Varid{s}\;\Varid{a}}.
Since there is no way to get hold of this unique label \ensuremath{\Varid{s}} in the rest
of the program, this guarantees that all \ensuremath{\Conid{STRef}}'s created by this
call of \ensuremath{\Varid{runST}} indeed point into the `heap' for which they were
handed out.

\begin{figure}
\begin{hscode}\SaveRestoreHook
\column{B}{@{}>{\hspre}l<{\hspost}@{}}%
\column{13}{@{}>{\hspre}l<{\hspost}@{}}%
\column{27}{@{}>{\hspre}c<{\hspost}@{}}%
\column{27E}{@{}l@{}}%
\column{31}{@{}>{\hspre}l<{\hspost}@{}}%
\column{34}{@{}>{\hspre}c<{\hspost}@{}}%
\column{34E}{@{}l@{}}%
\column{38}{@{}>{\hspre}l<{\hspost}@{}}%
\column{E}{@{}>{\hspre}l<{\hspost}@{}}%
\>[B]{}\Varid{newSTRef}{}\<[13]%
\>[13]{}\mathbin{::}{}\<[31]%
\>[31]{}\Varid{a}{}\<[34]%
\>[34]{}\to {}\<[34E]%
\>[38]{}\Conid{ST}\;\Varid{s}\;(\Conid{STRef}\;\Varid{s}\;\Varid{a}){}\<[E]%
\\
\>[B]{}\Varid{writeSTRef}{}\<[13]%
\>[13]{}\mathbin{::}\Conid{STRef}\;\Varid{s}\;\Varid{a}{}\<[27]%
\>[27]{}\to {}\<[27E]%
\>[31]{}\Varid{a}{}\<[34]%
\>[34]{}\to {}\<[34E]%
\>[38]{}\Conid{ST}\;\Varid{s}\;(){}\<[E]%
\\
\>[B]{}\Varid{readSTRef}{}\<[13]%
\>[13]{}\mathbin{::}\Conid{STRef}\;\Varid{s}\;\Varid{a}{}\<[34]%
\>[34]{}\to {}\<[34E]%
\>[38]{}\Conid{ST}\;\Varid{s}\;\Varid{a}{}\<[E]%
\\[\blanklineskip]%
\>[B]{}\Varid{runST}\mathbin{::}(\forall\;\Varid{s}\;.\;\Conid{ST}\;\Varid{s}\;\Varid{a})\to \Varid{a}{}\<[E]%
\ColumnHook
\end{hscode}\resethooks
\caption{Types of the ST operations}\label{fig:stcode}
\end{figure}
\begin{figure}
\begin{hscode}\SaveRestoreHook
\column{B}{@{}>{\hspre}l<{\hspost}@{}}%
\column{15}{@{}>{\hspre}l<{\hspost}@{}}%
\column{29}{@{}>{\hspre}l<{\hspost}@{}}%
\column{35}{@{}>{\hspre}l<{\hspost}@{}}%
\column{39}{@{}>{\hspre}l<{\hspost}@{}}%
\column{56}{@{}>{\hspre}l<{\hspost}@{}}%
\column{E}{@{}>{\hspre}l<{\hspost}@{}}%
\>[B]{}\mathbf{import}\;\Conid{\Conid{Control}.\Conid{Monad}.ST}{}\<[E]%
\\
\>[B]{}\mathbf{import}\;\Conid{\Conid{Data}.STRef}{}\<[E]%
\\[\blanklineskip]%
\>[B]{}\Varid{example}\mathrel{=}\mathbf{do}\;{}\<[15]%
\>[15]{}\Varid{r1}\leftarrow \Varid{newSTRef}\;\mathrm{2}{}\<[E]%
\\
\>[15]{}\Varid{v1}\leftarrow \Varid{readSTRef}\;\Varid{r1}{}\<[E]%
\\
\>[15]{}\Varid{r2}\leftarrow \Varid{newSTRef}\;(\Varid{show}\;\Varid{v1}){}\<[E]%
\\
\>[15]{}\Varid{modifySTRef}\;\Varid{r1}\;(\mathbin{+}\mathrm{3}){}\<[E]%
\\
\>[15]{}\Varid{v2}\leftarrow \Varid{readSTRef}\;\Varid{r2}{}\<[E]%
\\
\>[15]{}\mathbf{if}\;\Varid{v2}\equiv \text{\tt \char34 2\char34}\;{}\<[29]%
\>[29]{}\mathbf{then}\;{}\<[35]%
\>[35]{}\mathbf{do}\;{}\<[39]%
\>[39]{}\Varid{v1}\leftarrow \Varid{readSTRef}\;\Varid{r1}{}\<[E]%
\\
\>[39]{}\Varid{r3}\leftarrow \Varid{newSTRef}\;{}\<[56]%
\>[56]{}\Varid{v1}{}\<[E]%
\\
\>[39]{}\Varid{v3}\leftarrow \Varid{readSTRef}\;{}\<[56]%
\>[56]{}\Varid{r3}{}\<[E]%
\\
\>[39]{}\Varid{return}\;(\Varid{v2},\Varid{v1}){}\<[E]%
\\
\>[29]{}\mathbf{else}\;{}\<[35]%
\>[35]{}\Varid{return}\;(\text{\tt \char34 false\char34},\mathrm{7}){}\<[E]%
\\
\>[B]{}\Varid{demo}\mathrel{=}\Varid{runST}\;\Varid{example}{}\<[E]%
\ColumnHook
\end{hscode}\resethooks
\caption{An example of the \ensuremath{\Conid{ST}} monad}\label{fig:example}
\end{figure}
To show how to construct a state and how to read from it and write to it we give a small  \ensuremath{\Varid{example}} in Figure \ref{fig:example}; two variables \ensuremath{\Varid{r1}} and \ensuremath{\Varid{r2}} are introduced and initialized. In the \ensuremath{\Conid{True}} branch of the conditional expression we introduce another new variable. The value of \ensuremath{\Varid{demo}} evaluates to \ensuremath{(\text{\tt \char34 2\char34},\mathrm{5})}. 
This demonstrates that in general we will not be able to easily
determine from the program text 
how many variables will be created when running the code; we actually
have to mimic the execution. 
Note that the type of value held by each variable is fixed upon creating 
the variable, but not all variables hold a value of the same type. 

We present our implementation of the functions introduced above in two steps. In Section
\ref{state1} we show how a state may be  extended with new variables and how it is made accessible.
We will however not be able yet to access the variables.
In  Section \ref{state2} we extend the code and show  how we can create
references pointing into the state, and how to use these to read from and write to variables.

\subsection{Constructing a State}\label{state1}
\subsubsection{The type \ensuremath{\Conid{ST}}}
As we have seen in our example our state can accomodate many values of different types, 
so one of the first types which comes into mind to use for storing all 
these values is a nested Cartesian product. 
Indeed it is easy to do so if we know beforehand precisely which kind of values have to be stored and how many. 
If however the state evolves as a result of running the computation, 
and may even depend on values of variables introduced earlier this approach fails. 
Furthermore one of the distinguishing features of the \ensuremath{\Conid{ST}} monad is 
that the handed out references are labeled with a type, which uniquely identifies the state they point into; 
this is a property we definitely want to maintain since it is an important safety guarantee and 
we do not want to have dangling references into a state which has been garbage collected. 

In order to understand our solution it is helpful to try to forget about a specific evaluation order 
(as a reader might be naturally  inclined to do), and to move to a \emph{data-flow view} of our computations. 

\begin{figure}[t]

\begin{minipage}{0.45\columnwidth}

\begin{center}
\includegraphics[scale=0.5]{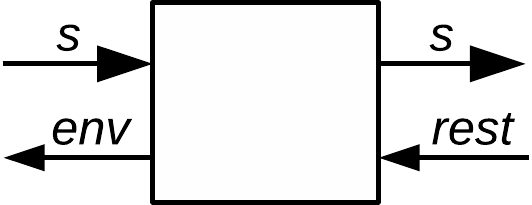}
\end{center}
\caption{\ensuremath{\Conid{ST}} type} \label{fig:envtype}

\end{minipage}\hfill
\begin{minipage}{0.45\columnwidth}
\begin{center}
\includegraphics[scale=0.45]{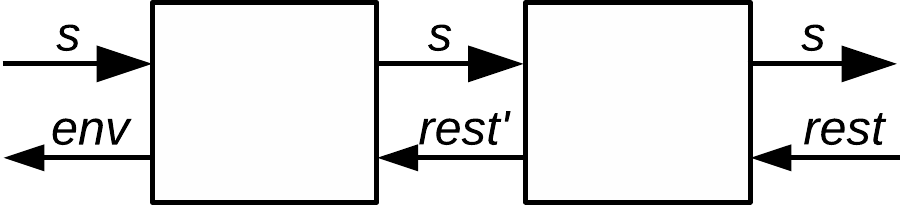}
\vspace{0.5mm}
\caption{Bind} \label{fig:bind}
\end{center}

\end{minipage}\hfill

\end{figure}

The box in Figure \ref{fig:envtype} represents a piece of stateful computation, i.e. a value of type \ensuremath{\Conid{ST}\;\Varid{s}\;\Varid{a}}, which is --for the time being-- a function type which:
\begin{itemize} 
\item takes two arguments of type \ensuremath{\Varid{s}} and \ensuremath{\Varid{rest}},
represented by incoming arrows 
\item returns a result of type \ensuremath{\Varid{s}}
and a result of type \ensuremath{\Varid{env}}, represented by outgoing arrows. These two
results will be represented by a tuple in our final code. 
\end{itemize} 

The Haskell type\footnote{This is the type as accepted by UHC. For GHC is is necessary to introduce an extra constructor when introducing the type \ensuremath{\Varid{env}}.} 
that corresponds to such a box is:
\begin{hscode}\SaveRestoreHook
\column{B}{@{}>{\hspre}l<{\hspost}@{}}%
\column{19}{@{}>{\hspre}l<{\hspost}@{}}%
\column{37}{@{}>{\hspre}l<{\hspost}@{}}%
\column{53}{@{}>{\hspre}l<{\hspost}@{}}%
\column{E}{@{}>{\hspre}l<{\hspost}@{}}%
\>[B]{}\mathbf{data}\;\Conid{ST}\;\Varid{s}\;\Varid{a}\mathrel{=}\Conid{ST}\;{}\<[19]%
\>[19]{}(\forall\;\Varid{t}_{\Varid{rest}}\;.\;{}\<[37]%
\>[37]{}\Varid{t}_{\Varid{rest}}\to \Varid{s}\to {}\<[53]%
\>[53]{}(\Varid{a},\exists\;\Varid{t}_{\Varid{env}}\;.\;\Varid{t}_{\Varid{env}},\Varid{s})){}\<[E]%
\ColumnHook
\end{hscode}\resethooks

The r\^ole of the various arrows is as follows:
\begin{itemize}
\item The input arrow \ensuremath{\Varid{rest}} represents the right-nested Cartesian
product of all the values that are possibly introduced by {\em
succeeding computations}.  Since our computation is indifferent to
this value its type is polymorphically quantified over this \ensuremath{\Varid{rest}}
type. Keep in mind that we deal with a language which has lazy
evaluation, so by the time this `value' is passed it will not have
been evaluated yet.
\item The result of type \ensuremath{\Varid{t}_{\Varid{env}}} again is a right-nested Cartesian
product, which has a value of the incoming type \ensuremath{\Varid{rest}} as its
tail.  This prefix to this tail corresponds to fields holding to the new variables
introduced by this \ensuremath{\Conid{ST}} value; hence it is an (possibly) extended version of
\ensuremath{\Varid{rest}}.  Since only the internals of our computation know how many
variables and of which types are added by this computation, we use an
existential result type.  Although some tail of the nested product
type \ensuremath{\Varid{t}_{\Varid{env}}} will be of the type \ensuremath{\Varid{t}_{\Varid{rest}}} this fact is not explicitly represented in the
\ensuremath{\Conid{ST}} type.
\item The remaining input and output arrows both carry a value of some
type \ensuremath{\Varid{s}}, which is the type of the final state of the overall
computation as ran by \ensuremath{\Varid{runST}}, eventually containing all created
variables.  This value is threaded though the computation, while the
variables contained in it are being read and written.  Since this value
has the shape of the final state this shape (and thus its type)
remains unchanged.  Whereas in the \ensuremath{\Conid{ST}} monad as built-in into GHC the
type of the state merely serves as a label indicating which `heap' we
are dealing with, here it is the actual heap in the form of the nested
Cartesian product that is being passed on.
\item In our pictures we have not shown the type \ensuremath{\Varid{a}}, since it does
not play a role in our explanation of how the state is constructed and
represented.
\end{itemize}

Statefull computations  can be composed into larger
computations by connecting their arrows as shown in \ref{fig:bind};
boxes to the left represent earlier computations and boxes to the
right succeeding computations.

\begin{figure}[t]
\begin{minipage}{0.45\columnwidth}

\begin{center}
\includegraphics[scale=0.45]{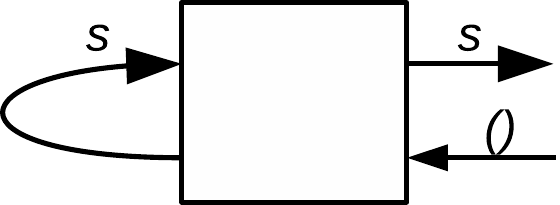}
\vspace{2mm}
\caption{Run} \label{fig:run}
\end{center}

\end{minipage}\hfill
\begin{minipage}{0.45\columnwidth}

\begin{center}
\includegraphics[scale=0.5]{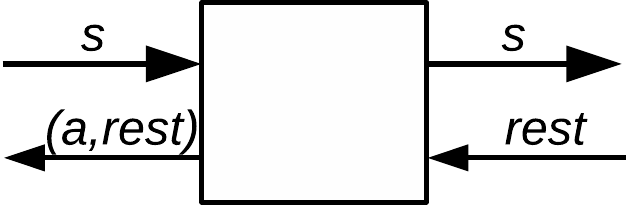}
\end{center}
\caption{\ensuremath{\Varid{insert}\;\Varid{a}}} \label{fig:insert}
\end{minipage}

\end{figure}

\subsubsection{\ensuremath{\Varid{runST}}} 
Before looking at the code of \ensuremath{\Varid{runST}} we take
a look at Figure~\ref{fig:run}.  When running the computation with the
function \ensuremath{\Varid{runST}}, the initial empty state \ensuremath{()} is passed as \ensuremath{\Varid{t}_{\Varid{rest}}}
argument at the very end of the computation. It emerges as the lazily
constructed value of some type \ensuremath{\Varid{t}_{\Varid{env}}} at the far left of our composed
boxes, containing all the variables (to be) introduced.  Now we decide
upon the type \ensuremath{\Varid{s}} and choose it to be the same as the returned type
\ensuremath{\Varid{t}_{\Varid{env}}}, and feed the value of type \ensuremath{\Varid{t}_{\Varid{env}}} back into the computation as
the \ensuremath{\Varid{s}} parameter.

In the code below we have given two definitions of runST: the first
one unannotated, and the second one containing 
explicit type annotations:

\begin{hscode}\SaveRestoreHook
\column{B}{@{}>{\hspre}l<{\hspost}@{}}%
\column{13}{@{}>{\hspre}l<{\hspost}@{}}%
\column{18}{@{}>{\hspre}l<{\hspost}@{}}%
\column{19}{@{}>{\hspre}l<{\hspost}@{}}%
\column{24}{@{}>{\hspre}l<{\hspost}@{}}%
\column{42}{@{}>{\hspre}l<{\hspost}@{}}%
\column{E}{@{}>{\hspre}l<{\hspost}@{}}%
\>[B]{}\Varid{runST}\mathbin{::}(\forall\;\Varid{s}\;.\;\Conid{ST}\;\Varid{s}\;\Varid{a})\to \Varid{a}{}\<[E]%
\\
\>[B]{}\Varid{runST}\;(\Conid{ST}\;\Varid{st})\mathrel{=}{}\<[19]%
\>[19]{}\mathbf{let}\;{}\<[24]%
\>[24]{}(\Varid{a},\Varid{env},\anonymous )\mathrel{=}\Varid{st}\;{}\<[42]%
\>[42]{}()\;\Varid{s}{}\<[E]%
\\
\>[19]{}\mathbf{in}\;{}\<[24]%
\>[24]{}\Varid{a}{}\<[E]%
\\
\>[B]{}\; {}\<[E]%
\\
\>[B]{}\Varid{runST}\mathbin{::}(\forall\;\Varid{s}\;.\;\Conid{ST}\;\Varid{s}\;\Varid{a})\to \Varid{a}{}\<[E]%
\\
\>[B]{}\Varid{runST}\;\Varid{st}\mathrel{=}{}\<[13]%
\>[13]{}\mathbf{let}\;{}\<[18]%
\>[18]{}(\Conid{ST}\;\Varid{st'})\mathrel{=}\Varid{st}\;[\mskip1.5mu \Varid{t}_{\Varid{env}}\mskip1.5mu]{}\<[E]%
\\
\>[18]{}(\Varid{a},\lessdot\Varid{t}_{\Varid{env}}, \Varid{env}\gtrdot,\anonymous )\mathrel{=}\Varid{st'}\;[\mskip1.5mu ()\mskip1.5mu]\;()\;\Varid{env}{}\<[E]%
\\
\>[13]{}\mathbf{in}\;{}\<[18]%
\>[18]{}\Varid{a}{}\<[E]%
\ColumnHook
\end{hscode}\resethooks

Focusing on the second definition we see that we instantiate the
polymorphic type of the value \ensuremath{\Varid{st}} with two types: the first type
parameter is the type \ensuremath{\Varid{t}_{\Varid{env}}} returned as the existential type by this
computation describing the lazily constructed complete state, and the
second is the type of the empty state \ensuremath{()}, which we feed in from the
far right end of our composed sequence of boxes.  We have omitted the
annotations for the type \ensuremath{\Varid{a}}, since they play no special r\^ole here.
Hence we do not only have a \ensuremath{\mathbf{letrec}} at the value-level, but also at
the type level: the type \ensuremath{\Varid{t}_{\Varid{env}}} which is `returned' as the type of
the existential \ensuremath{\Varid{env}}-part of the result is used as a type parameter
in a right hand side expression of the binding group!

Since the type of the final state \ensuremath{\Varid{s}} is universally quantified when
the computation is run, the user of the \ensuremath{\Conid{ST}} monad cannot assume
anything about it.  Thus, although we can easily add operators to
expose  the state \ensuremath{\Conid{ST}} as
we have defined thus far is completely useless. We can \ensuremath{\Varid{get}} a state,
and the only thing we can do with it is to \ensuremath{\Varid{put}} it back.
\begin{hscode}\SaveRestoreHook
\column{B}{@{}>{\hspre}l<{\hspost}@{}}%
\column{E}{@{}>{\hspre}l<{\hspost}@{}}%
\>[B]{}\Varid{get}\mathbin{::}\Conid{ST}\;\Varid{s}\;\Varid{s}{}\<[E]%
\\
\>[B]{}\Varid{get}\mathrel{=}\Conid{ST}\;(\lambda \Varid{env}\;\Varid{s}\to (\Varid{s},\Varid{env},\Varid{s})){}\<[E]%
\\
\>[B]{}\; {}\<[E]%
\\
\>[B]{}\Varid{put}\mathbin{::}\Varid{s}\to \Conid{ST}\;\Varid{s}\;(){}\<[E]%
\\
\>[B]{}\Varid{put}\;\Varid{s}\mathrel{=}\Conid{ST}\;(\lambda \Varid{env}\;\anonymous \to ((),\Varid{env},\Varid{s'})){}\<[E]%
\ColumnHook
\end{hscode}\resethooks

\subsubsection{\ensuremath{\Conid{Monad}}}
Computations are monads and can thus be composed with the monadic bind
(\ensuremath{\bind }) operator.
In Figure~\ref{fig:bind} we show how the state is constructed
from right to left and how this final state is modified in a
left-to-right traversal of the individual steps of the computation. 
%
\begin{hscode}\SaveRestoreHook
\column{B}{@{}>{\hspre}l<{\hspost}@{}}%
\column{3}{@{}>{\hspre}l<{\hspost}@{}}%
\column{5}{@{}>{\hspre}c<{\hspost}@{}}%
\column{5E}{@{}l@{}}%
\column{8}{@{}>{\hspre}l<{\hspost}@{}}%
\column{24}{@{}>{\hspre}l<{\hspost}@{}}%
\column{25}{@{}>{\hspre}l<{\hspost}@{}}%
\column{29}{@{}>{\hspre}l<{\hspost}@{}}%
\column{30}{@{}>{\hspre}l<{\hspost}@{}}%
\column{35}{@{}>{\hspre}l<{\hspost}@{}}%
\column{41}{@{}>{\hspre}c<{\hspost}@{}}%
\column{41E}{@{}l@{}}%
\column{43}{@{}>{\hspre}l<{\hspost}@{}}%
\column{48}{@{}>{\hspre}l<{\hspost}@{}}%
\column{49}{@{}>{\hspre}l<{\hspost}@{}}%
\column{57}{@{}>{\hspre}l<{\hspost}@{}}%
\column{58}{@{}>{\hspre}l<{\hspost}@{}}%
\column{65}{@{}>{\hspre}l<{\hspost}@{}}%
\column{E}{@{}>{\hspre}l<{\hspost}@{}}%
\>[B]{}\mathbf{instance}\;\Conid{Monad}\;(\Conid{ST}\;\Varid{s})\;\mathbf{where}{}\<[E]%
\\
\>[B]{}\hsindent{3}{}\<[3]%
\>[3]{}\Varid{return}\mathrel{=}\Varid{pure}{}\<[E]%
\\
\>[B]{}\hsindent{3}{}\<[3]%
\>[3]{}(\Conid{ST}\;\Varid{st}_{\Varid{a}})\bind \Varid{f}{}\<[E]%
\\
\>[3]{}\hsindent{2}{}\<[5]%
\>[5]{}\mathrel{=}{}\<[5E]%
\>[8]{}\Conid{ST}\;(\lambda \Varid{rest}\;\Varid{s}\to {}\<[24]%
\>[24]{}\mathbf{let}\;{}\<[29]%
\>[29]{}(\Varid{a},\Varid{env},\Varid{s'}){}\<[48]%
\>[48]{}\mathrel{=}\Varid{st}_{\Varid{a}}\;{}\<[57]%
\>[57]{}\Varid{rest'}\;\Varid{s})\;{}\<[E]%
\\
\>[29]{}\hsindent{1}{}\<[30]%
\>[30]{}(\Conid{ST}\;{}\<[35]%
\>[35]{}\Varid{st}_{\Varid{b}}{}\<[41]%
\>[41]{}){}\<[41E]%
\>[49]{}\mathrel{=}\Varid{f}\;{}\<[58]%
\>[58]{}\Varid{a}{}\<[E]%
\\
\>[29]{}\hsindent{1}{}\<[30]%
\>[30]{}(\Varid{b},{}\<[35]%
\>[35]{}\Varid{rest'},{}\<[43]%
\>[43]{}\Varid{s''}){}\<[49]%
\>[49]{}\mathrel{=}\Varid{st}_{\Varid{b}}\;{}\<[58]%
\>[58]{}\Varid{rest}\;{}\<[65]%
\>[65]{}\Varid{s'}{}\<[E]%
\\
\>[24]{}\hsindent{1}{}\<[25]%
\>[25]{}\mathbf{in}\;{}\<[30]%
\>[30]{}(\Varid{b},{}\<[35]%
\>[35]{}\Varid{env},{}\<[43]%
\>[43]{}\Varid{s''}){}\<[E]%
\ColumnHook
\end{hscode}\resethooks
Again this binder cannot be implemented with System-F existential types, because we pass \ensuremath{\Varid{rest'}}
returned by the second computation (\ensuremath{\Varid{st}_{\Varid{b}}}) to \ensuremath{\Varid{st}_{\Varid{a}}}, whereas the state of type \ensuremath{\Varid{s}} is passed in the other direction.

\subsubsection{\ensuremath{\Conid{Functor}} and \ensuremath{\Conid{Applicative}}}
In the last version of the Haskell libraries it is required that \ensuremath{\Conid{Monad}} instances are also \ensuremath{\Conid{Functor}} and  \ensuremath{\Conid{Applicative}} instances,
 amongst others to support the applicative \ensuremath{\mathbf{do}}-notation. 
So we define:
\begin{hscode}\SaveRestoreHook
\column{B}{@{}>{\hspre}l<{\hspost}@{}}%
\column{3}{@{}>{\hspre}l<{\hspost}@{}}%
\column{4}{@{}>{\hspre}l<{\hspost}@{}}%
\column{10}{@{}>{\hspre}l<{\hspost}@{}}%
\column{23}{@{}>{\hspre}l<{\hspost}@{}}%
\column{28}{@{}>{\hspre}l<{\hspost}@{}}%
\column{37}{@{}>{\hspre}l<{\hspost}@{}}%
\column{46}{@{}>{\hspre}l<{\hspost}@{}}%
\column{52}{@{}>{\hspre}l<{\hspost}@{}}%
\column{62}{@{}>{\hspre}l<{\hspost}@{}}%
\column{69}{@{}>{\hspre}l<{\hspost}@{}}%
\column{E}{@{}>{\hspre}l<{\hspost}@{}}%
\>[B]{}\mathbf{instance}\;\Conid{Functor}\;(\Conid{ST}\;\Varid{s})\;\mathbf{where}{}\<[E]%
\\
\>[B]{}\Varid{fmap}\;\Varid{f}\;(\Conid{ST}\;\Varid{st}){}\<[E]%
\\
\>[B]{}\hsindent{4}{}\<[4]%
\>[4]{}\mathrel{=}\Conid{ST}\;{}\<[10]%
\>[10]{}(\lambda \Varid{rest}\;\Varid{s}\to {}\<[23]%
\>[23]{}\mathbf{let}\;{}\<[28]%
\>[28]{}(\Varid{a},\Varid{rest'},\Varid{s'})\mathrel{=}\Varid{st}\;\Varid{rest}\;\Varid{s}{}\<[E]%
\\
\>[23]{}\mathbf{in}\;{}\<[28]%
\>[28]{}(\Varid{f}\;\Varid{a},\Varid{rest'},\Varid{s'}){}\<[E]%
\\
\>[10]{}){}\<[E]%
\\[\blanklineskip]%
\>[B]{}\mathbf{instance}\;\Conid{Applicative}\;(\Conid{ST}\;\Varid{s})\;\mathbf{where}{}\<[E]%
\\
\>[B]{}\hsindent{3}{}\<[3]%
\>[3]{}\Varid{pure}\;\Varid{a}\mathrel{=}\Conid{ST}\;(\lambda \Varid{rest}\;\Varid{s}\to {}\<[28]%
\>[28]{}(\Varid{a},\Varid{rest},\Varid{s})){}\<[E]%
\\
\>[B]{}\hsindent{3}{}\<[3]%
\>[3]{}(\Conid{ST}\;\Varid{st}_{\Varid{b2a}})\mathbin{{<\!\!\!*\!\!\!>}}\mathord{\sim}(\Conid{ST}\;\Varid{st}_{\Varid{b}}){}\<[E]%
\\
\>[3]{}\hsindent{1}{}\<[4]%
\>[4]{}\mathrel{=}\Conid{ST}\;{}\<[10]%
\>[10]{}(\lambda \Varid{rest}\;\Varid{s}\to {}\<[23]%
\>[23]{}\mathbf{let}\;{}\<[28]%
\>[28]{}(\Varid{b2a},{}\<[37]%
\>[37]{}\Varid{rest''},{}\<[46]%
\>[46]{}\Varid{s'}){}\<[52]%
\>[52]{}\mathrel{=}\Varid{st}_{\Varid{b2a}}\;{}\<[62]%
\>[62]{}\Varid{rest'}\;{}\<[69]%
\>[69]{}\Varid{s}{}\<[E]%
\\
\>[28]{}(\Varid{b},{}\<[37]%
\>[37]{}\Varid{rest'},{}\<[46]%
\>[46]{}\Varid{s''}){}\<[52]%
\>[52]{}\mathrel{=}\Varid{st}_{\Varid{b}}\;{}\<[62]%
\>[62]{}\Varid{rest}\;{}\<[69]%
\>[69]{}\Varid{s'}{}\<[E]%
\\
\>[23]{}\mathbf{in}\;{}\<[28]%
\>[28]{}(\Varid{b2a}\;\Varid{b},{}\<[37]%
\>[37]{}\Varid{rest''},{}\<[46]%
\>[46]{}\Varid{s''}){}\<[E]%
\\
\>[10]{}){}\<[E]%
\ColumnHook
\end{hscode}\resethooks
Note that we have used an irrefutable pattern for the right hand side parameter of \ensuremath{\mathbin{{<\!\!\!*\!\!\!>}}}; 
the evaluation of the right hand side should not be pushed any further than strictly necessary. 
It is important to note how the final state \ensuremath{\Varid{s}} is passed in a forward direction, 
whereas the future additions in \ensuremath{\Varid{rest}} are passed backwards through the computation; again we are tying a knot here.

\subsection{\ensuremath{\Varid{newSTRef}}}\label{state2}
\subsubsection{\ensuremath{\Varid{insert}}}
As a first step in showing how a state is constructed, we define the function \ensuremath{\Varid{insert}}, which takes a
value of some type \ensuremath{\Varid{a}} and extends the state with this value like a \ensuremath{\Varid{newSTRef}}, but it returns
the newly stored value instead of a reference to it. Again we have given the code with and without type annotations.

\begin{hscode}\SaveRestoreHook
\column{B}{@{}>{\hspre}l<{\hspost}@{}}%
\column{11}{@{}>{\hspre}l<{\hspost}@{}}%
\column{27}{@{}>{\hspre}l<{\hspost}@{}}%
\column{33}{@{}>{\hspre}l<{\hspost}@{}}%
\column{49}{@{}>{\hspre}c<{\hspost}@{}}%
\column{49E}{@{}l@{}}%
\column{52}{@{}>{\hspre}l<{\hspost}@{}}%
\column{E}{@{}>{\hspre}l<{\hspost}@{}}%
\>[B]{}\Varid{insert}\mathbin{::}\Varid{a}\to \Conid{ST}\;\Varid{s}\;\Varid{a}{}\<[E]%
\\
\>[B]{}\Varid{insert}\;\Varid{a}\mathrel{=}\Conid{ST}\;(\lambda \Varid{rest}\to {}\<[27]%
\>[27]{}\Varid{s}\to {}\<[33]%
\>[33]{}(\Varid{a},(\Varid{a},\Varid{rest}),\Varid{s})){}\<[E]%
\\
\>[B]{}\; {}\<[E]%
\\
\>[B]{}\Varid{insert}\mathrel{=}{}\<[11]%
\>[11]{}\Lambda\;\Varid{t}_{\Varid{a}}\to \Varid{a}\to \Conid{ST}\;(\Lambda\;\Varid{t}_{\Varid{s}}\;\Varid{t}_{\Varid{rest}}{}\<[E]%
\\
\>[11]{}\to \lambda (\Varid{rest}\mathbin{::}\Varid{t}_{\Varid{rest}})\;{}\<[33]%
\>[33]{}(\Varid{s}\mathbin{::}\Varid{t}_{\Varid{s}})\to {}\<[49]%
\>[49]{}({}\<[49E]%
\>[52]{}\Varid{a}\mathbin{::}\Varid{t}_{\Varid{a}}{}\<[E]%
\\
\>[49]{},{}\<[49E]%
\>[52]{}\lessdot(\Varid{t}_{\Varid{a}},\Varid{t}_{\Varid{rest}}), (\Varid{a},\Varid{rest})\gtrdot{}\<[E]%
\\
\>[49]{},{}\<[49E]%
\>[52]{}\Varid{s}\mathbin{::}\Varid{t}_{\Varid{s}}{}\<[E]%
\\
\>[49]{})){}\<[49E]%
\ColumnHook
\end{hscode}\resethooks
Figure~\ref{fig:insert} shows its effect on the state being
constructed. Notice  that the \ensuremath{\Varid{rest}} of the state needs no 
inspection in order to be able to extend it. Thus, due to lazy evaluation, even
infinite states can be constructed. Note furthermore that the function
\ensuremath{\Varid{insert}} should definitely not be made strict. The \ensuremath{\Varid{rest}} argument is likely
to depend on values read from a state of which the constructed \ensuremath{(\Varid{a},\Varid{rest})} pair is
a trailing component, and which is passed to it as the \ensuremath{\Varid{s}} parameter!

In the rest of this section we are going to extend our definitions such that we can create 
and use references into the constructed state.

\subsubsection{\ensuremath{\Conid{STRef}\;\Varid{s}\;\Varid{a}}}
Typed references over a nested Cartesian product are represented by 
the \ensuremath{\Conid{STRef}\;\Varid{s}\;\Varid{a}} GADT \cite{Sheard:2008:MBT:1346363.1346644,BSV09}, indexed by the type \ensuremath{\Varid{s}} of the Cartesian product representing our complete state
and \ensuremath{\Varid{a}} being the type of the  value referred to. The constructor \ensuremath{\Conid{RZ}} refers to the first element of the nested product,
whereas \ensuremath{\Conid{RS}} constructs the successor of an index \ensuremath{\Conid{STRef}\;\Varid{r}\;\Varid{a}} in a product of type \ensuremath{\Varid{r}}. The type \ensuremath{\Varid{b}} in this case is thus the first element which has to be skipped when indexing. 
\begin{hscode}\SaveRestoreHook
\column{B}{@{}>{\hspre}l<{\hspost}@{}}%
\column{4}{@{}>{\hspre}l<{\hspost}@{}}%
\column{E}{@{}>{\hspre}l<{\hspost}@{}}%
\>[B]{}\mathbf{data}\;\Conid{STRef}\;\Varid{s}\;\Varid{a}\;\mathbf{where}{}\<[E]%
\\
\>[B]{}\hsindent{4}{}\<[4]%
\>[4]{}\Conid{RZ}\mathbin{::}\Conid{STRef}\;(\Varid{a},\Varid{b})\;\Varid{a}{}\<[E]%
\\
\>[B]{}\hsindent{4}{}\<[4]%
\>[4]{}\Conid{RS}\mathbin{::}\Conid{STRef}\;\Varid{r}\;\Varid{a}\to \Conid{STRef}\;(\Varid{b},\Varid{r})\;\Varid{a}{}\<[E]%
\ColumnHook
\end{hscode}\resethooks

With such references, look-ups and modifications can be performed safely. 
The type system makes sure that no pointers can point outside of the 
structure and that the type pointed to is the type we expect: 
\begin{hscode}\SaveRestoreHook
\column{B}{@{}>{\hspre}l<{\hspost}@{}}%
\column{10}{@{}>{\hspre}l<{\hspost}@{}}%
\column{13}{@{}>{\hspre}l<{\hspost}@{}}%
\column{18}{@{}>{\hspre}l<{\hspost}@{}}%
\column{21}{@{}>{\hspre}l<{\hspost}@{}}%
\column{25}{@{}>{\hspre}l<{\hspost}@{}}%
\column{28}{@{}>{\hspre}l<{\hspost}@{}}%
\column{E}{@{}>{\hspre}l<{\hspost}@{}}%
\>[B]{}\Varid{rlookup}\mathbin{::}\Conid{STRef}\;\Varid{s}\;\Varid{a}\to \Varid{s}\to \Varid{a}{}\<[E]%
\\
\>[B]{}\Varid{rlookup}\;{}\<[10]%
\>[10]{}\Conid{RZ}\;{}\<[18]%
\>[18]{}(\Varid{a},\anonymous ){}\<[25]%
\>[25]{}\mathrel{=}\Varid{a}{}\<[E]%
\\
\>[B]{}\Varid{rlookup}\;{}\<[10]%
\>[10]{}(\Conid{RS}\;\Varid{r})\;{}\<[18]%
\>[18]{}(\anonymous ,\Varid{b}){}\<[25]%
\>[25]{}\mathrel{=}\Varid{rlookup}\;\Varid{r}\;\Varid{b}{}\<[E]%
\\[\blanklineskip]%
\>[B]{}\Varid{rmodify}\mathbin{::}(\Varid{a}\to \Varid{a})\to \Conid{STRef}\;\Varid{s}\;\Varid{a}\to \Varid{s}\to \Varid{s}{}\<[E]%
\\
\>[B]{}\Varid{rmodify}\;{}\<[10]%
\>[10]{}\Varid{f}\;{}\<[13]%
\>[13]{}\Conid{RZ}\;{}\<[21]%
\>[21]{}(\Varid{a},\Varid{r}){}\<[28]%
\>[28]{}\mathrel{=}(\Varid{f}\;\Varid{a},\Varid{r}){}\<[E]%
\\
\>[B]{}\Varid{rmodify}\;{}\<[10]%
\>[10]{}\Varid{f}\;{}\<[13]%
\>[13]{}(\Conid{RS}\;\Varid{r})\;{}\<[21]%
\>[21]{}(\Varid{a},\Varid{b}){}\<[28]%
\>[28]{}\mathrel{=}(\Varid{a},\Varid{rmodify}\;\Varid{f}\;\Varid{r}\;\Varid{b}){}\<[E]%
\ColumnHook
\end{hscode}\resethooks

\begin{figure}[t]

\begin{center}
\includegraphics[scale=0.4]{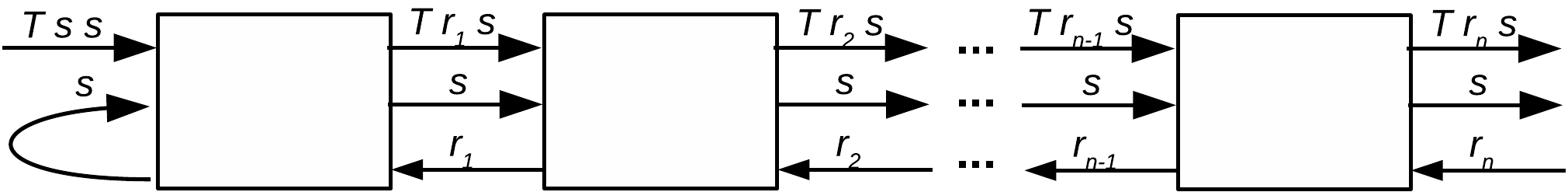}
\caption{\ensuremath{\Conid{ST}} type with references} \label{fig:final}
\end{center}

\end{figure}

\subsubsection{Transforming \ensuremath{\Conid{STRef}}'s}
When adding a new location to the state, with the function \ensuremath{\Varid{newSTRef}},
the reference to its position in the final state is to be returned.
\begin{hscode}\SaveRestoreHook
\column{B}{@{}>{\hspre}l<{\hspost}@{}}%
\column{3}{@{}>{\hspre}l<{\hspost}@{}}%
\column{E}{@{}>{\hspre}l<{\hspost}@{}}%
\>[3]{}\Varid{newSTRef}\mathbin{::}\Varid{a}\to \Conid{ST}\;\Varid{s}\;(\Conid{STRef}\;\Varid{s}\;\Varid{a}){}\<[E]%
\ColumnHook
\end{hscode}\resethooks
%
If we look at the \ensuremath{\Conid{ST}} type defined in the previous subsection (and represented in Figure~\ref{fig:envtype}),
there is no way to relate the state \ensuremath{\Varid{env}} constructed by this computation and the final state.
Thus, we have to extend the \ensuremath{\Conid{ST}} type so this relation becomes available.

For this purpose we define a type \ensuremath{\Conid{T}\;s_1\;s_2} which represents a  transformation of a reference into a structure
\ensuremath{s_1} to one in  \ensuremath{s_2}.
\begin{hscode}\SaveRestoreHook
\column{B}{@{}>{\hspre}l<{\hspost}@{}}%
\column{3}{@{}>{\hspre}l<{\hspost}@{}}%
\column{E}{@{}>{\hspre}l<{\hspost}@{}}%
\>[B]{}\mathbf{newtype}\;\Conid{T}\;s_1\;s_2{}\<[E]%
\\
\>[B]{}\hsindent{3}{}\<[3]%
\>[3]{}\mathrel{=}\Conid{T}\;\{\mskip1.5mu \Varid{unT}\mathbin{::}\forall\;\Varid{a}\;.\;\Conid{STRef}\;s_1\;\Varid{a}\to \Conid{STRef}\;s_2\;\Varid{a}\mskip1.5mu\}{}\<[E]%
\ColumnHook
\end{hscode}\resethooks
%
\subsubsection{Our extended version of \ensuremath{\Conid{ST}}}
We now change the type \ensuremath{\Conid{ST}} such that transformations from references in the \ensuremath{\Varid{rest}} state to references in the final state
are passed on from left to right, together with the state of type \ensuremath{\Varid{s}}. One may think of it as a counter keeping track of 
how many elements have thus far been added to the state. The counter 
is indexed with the type which remains of the final state provided the counted number of elements have been removed.
Note that because this type is indexed by the type of \ensuremath{\Varid{env}} it is actually determined by the passed value \ensuremath{\Varid{rest}} 
together with the number and types of the new variables added by this computation. 
\begin{hscode}\SaveRestoreHook
\column{B}{@{}>{\hspre}l<{\hspost}@{}}%
\column{3}{@{}>{\hspre}l<{\hspost}@{}}%
\column{9}{@{}>{\hspre}c<{\hspost}@{}}%
\column{9E}{@{}l@{}}%
\column{10}{@{}>{\hspre}c<{\hspost}@{}}%
\column{10E}{@{}l@{}}%
\column{14}{@{}>{\hspre}l<{\hspost}@{}}%
\column{E}{@{}>{\hspre}l<{\hspost}@{}}%
\>[B]{}\mathbf{data}\;\Conid{ST}\;\Varid{s}\;\Varid{a}{}\<[E]%
\\
\>[B]{}\hsindent{3}{}\<[3]%
\>[3]{}\mathrel{=}\Conid{ST}\;{}\<[9]%
\>[9]{}({}\<[9E]%
\>[14]{}\forall\;\Varid{t}_{\Varid{rest}}\;.\;\Varid{t}_{\Varid{rest}}\to \Varid{s}{}\<[E]%
\\
\>[9]{}\hsindent{1}{}\<[10]%
\>[10]{}\to {}\<[10E]%
\>[14]{}{\bar{\exists}}\;\Varid{t}_{\Varid{env}}\;.\;\Conid{T}\;\Varid{t}_{\Varid{env}}\;\Varid{s}\to (\Varid{a},\Varid{t}_{\Varid{env}},\Varid{s},\Conid{T}\;\Varid{t}_{\Varid{rest}}\;\Varid{s})){}\<[E]%
\ColumnHook
\end{hscode}\resethooks

Notice that we are using our new
quantifier \ensuremath{{\bar{\exists}}}, which is to be interpreted as follows: as before 
we think of the type \ensuremath{\Varid{env}}
as being determined by the call to the function and made available as part of the result. Since we want to make use
of this type in specifying the type of one of the parameters we have
somehow to extended the scope `forward`. 

The instances given  have now to be extended to pass these transformations through the constituting computations. 
We do so for the \ensuremath{\Conid{Monad}} instance. The others follow trivially.
\begin{hscode}\SaveRestoreHook
\column{B}{@{}>{\hspre}l<{\hspost}@{}}%
\column{3}{@{}>{\hspre}l<{\hspost}@{}}%
\column{5}{@{}>{\hspre}l<{\hspost}@{}}%
\column{8}{@{}>{\hspre}l<{\hspost}@{}}%
\column{12}{@{}>{\hspre}c<{\hspost}@{}}%
\column{12E}{@{}l@{}}%
\column{16}{@{}>{\hspre}l<{\hspost}@{}}%
\column{21}{@{}>{\hspre}l<{\hspost}@{}}%
\column{26}{@{}>{\hspre}l<{\hspost}@{}}%
\column{34}{@{}>{\hspre}l<{\hspost}@{}}%
\column{40}{@{}>{\hspre}l<{\hspost}@{}}%
\column{51}{@{}>{\hspre}l<{\hspost}@{}}%
\column{59}{@{}>{\hspre}l<{\hspost}@{}}%
\column{66}{@{}>{\hspre}l<{\hspost}@{}}%
\column{70}{@{}>{\hspre}l<{\hspost}@{}}%
\column{E}{@{}>{\hspre}l<{\hspost}@{}}%
\>[B]{}\mathbf{instance}\;\Conid{Monad}\;(\Conid{ST}\;\Varid{s})\;\mathbf{where}{}\<[E]%
\\
\>[B]{}\hsindent{3}{}\<[3]%
\>[3]{}\Varid{return}\;\Varid{a}\mathrel{=}\Conid{ST}\;(\lambda \Varid{env}\;\Varid{s}\;tr_{env \mapsto s}\to (\Varid{a},\Varid{env},\Varid{s},tr_{env \mapsto s})){}\<[E]%
\\[\blanklineskip]%
\>[B]{}\hsindent{3}{}\<[3]%
\>[3]{}(\Conid{ST}\;{}\<[8]%
\>[8]{}\Varid{st}_{\Varid{a}})\bind \Varid{f}{}\<[E]%
\\
\>[3]{}\hsindent{2}{}\<[5]%
\>[5]{}\mathrel{=}\Conid{ST}\mathbin{\$}\lambda \Varid{rest}\;\Varid{s}\;tr_{env \mapsto s}{}\<[E]%
\\
\>[5]{}\hsindent{7}{}\<[12]%
\>[12]{}\to {}\<[12E]%
\>[16]{}\mathbf{let}\;{}\<[21]%
\>[21]{}(\Varid{a},{}\<[26]%
\>[26]{}\Varid{env},{}\<[34]%
\>[34]{}\Varid{s'},{}\<[40]%
\>[40]{}tr_{rest' \mapsto s}){}\<[51]%
\>[51]{}\mathrel{=}\Varid{st}_{\Varid{a}}\;{}\<[59]%
\>[59]{}\Varid{rest'}\;{}\<[66]%
\>[66]{}\Varid{s}\;{}\<[70]%
\>[70]{}tr_{env \mapsto s}{}\<[E]%
\\
\>[21]{}(\Conid{ST}\;{}\<[26]%
\>[26]{}\Varid{st}_{\Varid{b}}){}\<[51]%
\>[51]{}\mathrel{=}\Varid{f}\;\Varid{a}{}\<[E]%
\\
\>[21]{}(\Varid{b},{}\<[26]%
\>[26]{}\Varid{rest'},{}\<[34]%
\>[34]{}\Varid{s''},{}\<[40]%
\>[40]{}tr_{rest \mapsto s}){}\<[51]%
\>[51]{}\mathrel{=}\Varid{st}_{\Varid{b}}\;{}\<[59]%
\>[59]{}\Varid{rest}\;{}\<[66]%
\>[66]{}\Varid{s'}\;{}\<[70]%
\>[70]{}tr_{rest' \mapsto s}{}\<[E]%
\\
\>[16]{}\mathbf{in}\;{}\<[21]%
\>[21]{}(\Varid{b},{}\<[26]%
\>[26]{}\Varid{env},{}\<[34]%
\>[34]{}\Varid{s''},{}\<[40]%
\>[40]{}tr_{rest \mapsto s}){}\<[E]%
\ColumnHook
\end{hscode}\resethooks

In Figure~\ref{fig:final} we show the types involved when running a composition of computations.
For the leftmost computation, the transformation is just the identity function,
since this is the type of the final state (no more locations will be added). Note that because we have again chosen the returned \ensuremath{\Varid{env}} to be the value to pass on as the initial state \ensuremath{\Varid{s}} these have the same type, and thus \ensuremath{\Conid{T}\;\Varid{id}\mathbin{::}\Conid{T}\;\Varid{t}_{\Varid{env}}\;\Varid{tenv}} has the correct type.

\begin{hscode}\SaveRestoreHook
\column{B}{@{}>{\hspre}l<{\hspost}@{}}%
\column{3}{@{}>{\hspre}l<{\hspost}@{}}%
\column{8}{@{}>{\hspre}l<{\hspost}@{}}%
\column{19}{@{}>{\hspre}l<{\hspost}@{}}%
\column{59}{@{}>{\hspre}l<{\hspost}@{}}%
\column{E}{@{}>{\hspre}l<{\hspost}@{}}%
\>[B]{}\Varid{runST}\mathbin{::}(\forall\;{}\<[19]%
\>[19]{}\Varid{s}\;.\;\Conid{ST}\;\Varid{s}\;\Varid{a})\to \Varid{a}{}\<[E]%
\\
\>[B]{}\Varid{runST}\;(\Conid{ST}\;\Varid{st})\mathrel{=}{}\<[E]%
\\
\>[B]{}\hsindent{3}{}\<[3]%
\>[3]{}\mathbf{let}\;{}\<[8]%
\>[8]{}(\Varid{a},\lessdot\Varid{t}_{\Varid{env}}, \Varid{env}\gtrdot,\anonymous ,\anonymous )\mathrel{=}\Varid{st}\;()\;\Varid{env}\;(\Conid{T}\;\Varid{id})\;\mathbf{in}\;{}\<[59]%
\>[59]{}\Varid{a}{}\<[E]%
\ColumnHook
\end{hscode}\resethooks

When adding a new location, the reference to this new location in the final state
is obtained by applying the transformation \ensuremath{tr_{env \mapsto s}} (with type \ensuremath{\Conid{T}\;\Varid{env}\;\Varid{s}}) to
a reference to the first position in the just extended state.
The transformation of references for the succeeding computations 
is obtained by composing current transformation with \ensuremath{\Conid{RS}};
i.e. references in the rest of the state point to locations in the second component of the pair \ensuremath{(\Varid{a},\Varid{rest})}.

\begin{hscode}\SaveRestoreHook
\column{B}{@{}>{\hspre}l<{\hspost}@{}}%
\column{20}{@{}>{\hspre}l<{\hspost}@{}}%
\column{24}{@{}>{\hspre}l<{\hspost}@{}}%
\column{E}{@{}>{\hspre}l<{\hspost}@{}}%
\>[B]{}\Varid{newSTRef}\mathbin{::}\Varid{a}\to \Conid{ST}\;\Varid{s}\;(\Conid{STRef}\;\Varid{s}\;\Varid{a}){}\<[E]%
\\
\>[B]{}\Varid{newSTRef}\;\Varid{a}\mathrel{=}\Conid{ST}\mathbin{\$}{}\<[20]%
\>[20]{}\lambda \Varid{rest}\;\Varid{s}\;tr_{env \mapsto s}{}\<[E]%
\\
\>[20]{}\to {}\<[24]%
\>[24]{}((\Varid{unT}\;tr_{env \mapsto s})\;\Conid{RZ}{}\<[E]%
\\
\>[24]{},(\Varid{a},\Varid{rest}){}\<[E]%
\\
\>[24]{},\Varid{s}{}\<[E]%
\\
\>[24]{},\Conid{T}\;(\Varid{unT}\;tr_{env \mapsto s}\;.\;\Conid{RS}){}\<[E]%
\\
\>[24]{}){}\<[E]%
\ColumnHook
\end{hscode}\resethooks

Having a reference to an element in the final state, 
to obtain the referred value is to perform an \ensuremath{\Varid{rlookup}} in the state \ensuremath{\Varid{s}}
(traveling from left-to-right).
\begin{hscode}\SaveRestoreHook
\column{B}{@{}>{\hspre}l<{\hspost}@{}}%
\column{3}{@{}>{\hspre}l<{\hspost}@{}}%
\column{11}{@{}>{\hspre}l<{\hspost}@{}}%
\column{E}{@{}>{\hspre}l<{\hspost}@{}}%
\>[B]{}\Varid{readSTRef}\mathbin{::}\Conid{STRef}\;\Varid{s}\;\Varid{a}\to \Conid{ST}\;\Varid{s}\;\Varid{a}{}\<[E]%
\\
\>[B]{}\Varid{readSTRef}\;\Varid{r}{}\<[E]%
\\
\>[B]{}\hsindent{3}{}\<[3]%
\>[3]{}\mathrel{=}\Conid{ST}\mathbin{\$}{}\<[11]%
\>[11]{}\lambda \Varid{env}\;\Varid{s}\;tr_{env \mapsto s}\to (\Varid{rlookup}\;\Varid{r}\;\Varid{s},\Varid{env},\Varid{s},tr_{env \mapsto s}){}\<[E]%
\ColumnHook
\end{hscode}\resethooks
%
Similarly, we can overwrite or modify stored values.
\begin{hscode}\SaveRestoreHook
\column{B}{@{}>{\hspre}l<{\hspost}@{}}%
\column{3}{@{}>{\hspre}l<{\hspost}@{}}%
\column{11}{@{}>{\hspre}l<{\hspost}@{}}%
\column{29}{@{}>{\hspre}l<{\hspost}@{}}%
\column{E}{@{}>{\hspre}l<{\hspost}@{}}%
\>[B]{}\Varid{writeSTRef}\mathbin{::}\Conid{STRef}\;\Varid{s}\;\Varid{a}\to \Varid{a}\to \Conid{ST}\;\Varid{s}\;(){}\<[E]%
\\
\>[B]{}\Varid{writeSTRef}\;\Varid{r}\;\Varid{a}{}\<[E]%
\\
\>[B]{}\hsindent{3}{}\<[3]%
\>[3]{}\mathrel{=}\Conid{ST}\mathbin{\$}{}\<[11]%
\>[11]{}(\lambda \Varid{env}\;\Varid{s}\;tr_{env \mapsto s}\to {}\<[29]%
\>[29]{}((){}\<[E]%
\\
\>[29]{},\Varid{env}{}\<[E]%
\\
\>[29]{},\Varid{rmodify}\;(\Varid{const}\;\Varid{a})\;\Varid{r}\;\Varid{s}{}\<[E]%
\\
\>[29]{},tr_{env \mapsto s}{}\<[E]%
\\
\>[29]{})){}\<[E]%
\\[\blanklineskip]%
\>[B]{}\Varid{modifySTRef}\mathbin{::}\Conid{STRef}\;\Varid{s}\;\Varid{a}\to (\Varid{a}\to \Varid{a})\to \Conid{ST}\;\Varid{s}\;(){}\<[E]%
\\
\>[B]{}\Varid{modifySTRef}\;\Varid{r}\;\Varid{f}{}\<[E]%
\\
\>[B]{}\hsindent{3}{}\<[3]%
\>[3]{}\mathrel{=}\Conid{ST}\mathbin{\$}{}\<[11]%
\>[11]{}\lambda \Varid{env}\;\Varid{s}\;tr_{env \mapsto s}\to ((),\Varid{env},\Varid{rmodify}\;\Varid{f}\;\Varid{r}\;\Varid{s},tr_{env \mapsto s}){}\<[E]%
\ColumnHook
\end{hscode}\resethooks

A nice property of our representation of pointers is that they may be
compared, and if two pointers are found to be equal then the 
GADT-based type system returns a proof that the type of the value pointed at is
the same.

\section{Fine tuning}
\subsection{Do we need \ensuremath{\exists} and \ensuremath{{\bar{\exists}}}?}
When we compare the type rules for \ensuremath{\exists} and \ensuremath{{\bar{\exists}}} we see that
they only differ for function types. One may argue that the case for having
a normal \ensuremath{\exists} does not make much sense for a function type; so
when we decide to treat the \ensuremath{\exists} for a function as an \ensuremath{{\bar{\exists}}}
there is no need for an extra symbol. 

We do not loose expressivity, since 
if we really want to have the classical existentially quantified function,
of which the unpacked version can be called at 
multiple places  we can easily tuple it with a dummy value into an
existential pair:

\begin{hscode}\SaveRestoreHook
\column{B}{@{}>{\hspre}l<{\hspost}@{}}%
\column{6}{@{}>{\hspre}l<{\hspost}@{}}%
\column{16}{@{}>{\hspre}l<{\hspost}@{}}%
\column{23}{@{}>{\hspre}l<{\hspost}@{}}%
\column{E}{@{}>{\hspre}l<{\hspost}@{}}%
\>[B]{}\Varid{d\char95 f}\mathbin{::}\exists\;\Varid{x}\;.\;(\Varid{x},\Varid{x}\to \Conid{Bool}\to (\Varid{x},\Conid{Int})\mathrel{=}(\mathrm{3},\lambda \Varid{v}\;\anonymous \to (\Varid{v}\mathbin{+}\mathrm{1},\Varid{v})){}\<[E]%
\\
\>[B]{}\mathbf{let}\;{}\<[6]%
\>[6]{}\lessdot\Varid{t}, (\anonymous ,\Varid{g})\gtrdot{}\<[23]%
\>[23]{}\mathrel{=}\Varid{d\char95 f}\mbox{\onelinecomment  unpacking}{}\<[E]%
\\
\>[6]{}(\Varid{i1},\Varid{v1}){}\<[16]%
\>[16]{}\mathrel{=}\Varid{g}\;\Varid{i2}\;\Conid{True}{}\<[E]%
\\
\>[6]{}(\Varid{i2},\Varid{v2}){}\<[16]%
\>[16]{}\mathrel{=}\Varid{g}\;\Varid{i1}\;\Conid{False}{}\<[E]%
\\
\>[B]{}\mathbf{in}\;\Varid{v1}\mathbin{...}{}\<[E]%
\ColumnHook
\end{hscode}\resethooks

Note that in \ensuremath{\Varid{d\char95 f}} the  choice for type \ensuremath{\Varid{x}} can no longer depend on the passed \ensuremath{\Conid{Bool}} argument.

\subsection{Extending the class system}
An indispensable component of the Haskell type system is its class
system, which makes it possible to pass extra information about a
polymorphic value to a function. In a similar way we may want to
provide extra information to the calling context. 
Currently Haskell only allows constructors to be constrained by classes
if these classes refer to existential types. We propose to generalise
this: a constraint on a constructor just packs an extra dictionary in
the record, and pattern matching on such a constructor brings the
class instance in scope. 
With this  extension we can rewrite our solution for the sorting tree to the code 
given in Figure~\ref{fig:withclasses}.

Note how, just as in the case with polymorphic functions, the class
instances will be automatically constructed, passed around and
accessed.

\begin{figure}
\begin{hscode}\SaveRestoreHook
\column{B}{@{}>{\hspre}l<{\hspost}@{}}%
\column{3}{@{}>{\hspre}l<{\hspost}@{}}%
\column{6}{@{}>{\hspre}l<{\hspost}@{}}%
\column{11}{@{}>{\hspre}l<{\hspost}@{}}%
\column{13}{@{}>{\hspre}l<{\hspost}@{}}%
\column{16}{@{}>{\hspre}l<{\hspost}@{}}%
\column{22}{@{}>{\hspre}c<{\hspost}@{}}%
\column{22E}{@{}l@{}}%
\column{23}{@{}>{\hspre}l<{\hspost}@{}}%
\column{25}{@{}>{\hspre}l<{\hspost}@{}}%
\column{29}{@{}>{\hspre}l<{\hspost}@{}}%
\column{35}{@{}>{\hspre}l<{\hspost}@{}}%
\column{39}{@{}>{\hspre}l<{\hspost}@{}}%
\column{41}{@{}>{\hspre}l<{\hspost}@{}}%
\column{43}{@{}>{\hspre}l<{\hspost}@{}}%
\column{46}{@{}>{\hspre}l<{\hspost}@{}}%
\column{E}{@{}>{\hspre}l<{\hspost}@{}}%
\>[B]{}\mathbf{class}\;\Conid{Insertable}\;\Varid{cl}\;\mathbf{where}{}\<[E]%
\\
\>[B]{}\hsindent{3}{}\<[3]%
\>[3]{}\Varid{insert}\mathbin{::}\Conid{Int}\to \Varid{cl}\to (\Conid{Int},\Varid{cl}){}\<[E]%
\\[\blanklineskip]%
\>[B]{}\mathbf{instance}\;\Conid{Insertable}\;\Varid{cl}\Rightarrow \Conid{Insertable}\;(\Conid{Int},\Varid{cl})\;\mathbf{where}{}\<[E]%
\\
\>[B]{}\hsindent{3}{}\<[3]%
\>[3]{}\Varid{insert}\;\Varid{w}\;{}\<[13]%
\>[13]{}(\Varid{x},\Varid{xs}){}\<[22]%
\>[22]{}\mathrel{=}{}\<[22E]%
\>[25]{}\mathbf{if}\;\Varid{w}\mathbin{<}\Varid{x}\;{}\<[35]%
\>[35]{}\mathbf{then}\;{}\<[41]%
\>[41]{}(\Varid{w},(\Varid{x},\Varid{xs})){}\<[E]%
\\
\>[35]{}\mathbf{else}\;{}\<[41]%
\>[41]{}(\Varid{x},\Varid{insert}\;\Varid{w}\;\Varid{xs})){}\<[E]%
\\[\blanklineskip]%
\>[B]{}\mathbf{instance}\;\Conid{Insertable}\;()\;\mathbf{where}{}\<[E]%
\\
\>[B]{}\hsindent{3}{}\<[3]%
\>[3]{}\Varid{insert}\;\Varid{w}\;{}\<[13]%
\>[13]{}(){}\<[22]%
\>[22]{}\mathrel{=}{}\<[22E]%
\>[25]{}(\Varid{w},()){}\<[E]%
\\[\blanklineskip]%
\>[B]{}\mathbf{data}\;\Conid{OrdList}\;\Varid{cl}\;\mathbf{where}{}\<[E]%
\\
\>[B]{}\hsindent{3}{}\<[3]%
\>[3]{}\Conid{OrdList}\mathbin{::}\Conid{Insertable}\;\Varid{cl}\Rightarrow \Varid{cl}\to \Conid{OrdList}\;{}\<[46]%
\>[46]{}\Varid{cl}{}\<[E]%
\\[\blanklineskip]%
\>[B]{}\Varid{sortTree''}\;(\Conid{Leaf}\;\Varid{v})\;{}\<[23]%
\>[23]{}(\Conid{OrdList}\;\Varid{rest}){}\<[39]%
\>[39]{}\mathord{\sim}(\Varid{x},\Varid{xs}){}\<[E]%
\\
\>[B]{}\hsindent{3}{}\<[3]%
\>[3]{}\mathrel{=}(\Conid{OrdList}\;{}\<[16]%
\>[16]{}(\Varid{insert}\;\Varid{v}\;\Varid{rest}),\Varid{xs},\Conid{Leaf}\;\Varid{x}){}\<[E]%
\\[\blanklineskip]%
\>[B]{}\Varid{sortTree''}\;(\Conid{Bin}\;\Varid{l}\;\Varid{r})\;\Varid{rest}\;{}\<[29]%
\>[29]{}\Varid{xs}{}\<[E]%
\\
\>[B]{}\hsindent{3}{}\<[3]%
\>[3]{}\mathrel{=}{}\<[6]%
\>[6]{}\mathbf{let}\;{}\<[11]%
\>[11]{}(\Varid{vl},\Varid{xsl},\Varid{tl}){}\<[29]%
\>[29]{}\mathrel{=}\Varid{sortTree''}\;{}\<[43]%
\>[43]{}\Varid{l}\;\Varid{vr}\;\Varid{xs}{}\<[E]%
\\
\>[11]{}(\Varid{vr},\Varid{xsr},\Varid{tr}){}\<[29]%
\>[29]{}\mathrel{=}\Varid{sortTree''}\;{}\<[43]%
\>[43]{}\Varid{r}\;\Varid{rest}\;\Varid{xsl}{}\<[E]%
\\
\>[6]{}\mathbf{in}\;{}\<[11]%
\>[11]{}(\Varid{vl},\Varid{xsr},\Conid{Bin}\;\Varid{tl}\;\Varid{tr}){}\<[E]%
\ColumnHook
\end{hscode}\resethooks
\caption{Using classes}\label{fig:withclasses}
\end{figure}

As a final extension we show how the new extension comes in handy in the case of the use of guards. 
Suppose we want to change our tree sorting algorithm such that we sort 
the sublists containing the even and the odd leaf values separately. 
This can be done by duplicating the parameters:
\begin{hscode}\SaveRestoreHook
\column{B}{@{}>{\hspre}l<{\hspost}@{}}%
\column{3}{@{}>{\hspre}l<{\hspost}@{}}%
\column{4}{@{}>{\hspre}l<{\hspost}@{}}%
\column{9}{@{}>{\hspre}l<{\hspost}@{}}%
\column{11}{@{}>{\hspre}l<{\hspost}@{}}%
\column{13}{@{}>{\hspre}l<{\hspost}@{}}%
\column{15}{@{}>{\hspre}l<{\hspost}@{}}%
\column{24}{@{}>{\hspre}l<{\hspost}@{}}%
\column{27}{@{}>{\hspre}l<{\hspost}@{}}%
\column{28}{@{}>{\hspre}l<{\hspost}@{}}%
\column{31}{@{}>{\hspre}l<{\hspost}@{}}%
\column{38}{@{}>{\hspre}l<{\hspost}@{}}%
\column{45}{@{}>{\hspre}l<{\hspost}@{}}%
\column{47}{@{}>{\hspre}l<{\hspost}@{}}%
\column{58}{@{}>{\hspre}l<{\hspost}@{}}%
\column{E}{@{}>{\hspre}l<{\hspost}@{}}%
\>[B]{}\Varid{sortTree''}\;{}\<[13]%
\>[13]{}(\Conid{Leaf}\;\Varid{v})\;{}\<[24]%
\>[24]{}((\Conid{OrdList}\;\Varid{e},{}\<[38]%
\>[38]{}\Varid{o})\; {}\<[45]%
\>[45]{}\mathord{\sim}((\Varid{x},\Varid{xs}),\Varid{ys}){}\<[E]%
\\
\>[B]{}\hsindent{3}{}\<[3]%
\>[3]{}\mid \Varid{even}\;{}\<[11]%
\>[11]{}\Varid{v}{}\<[15]%
\>[15]{}\mathrel{=}((\Conid{OrdList}\;{}\<[28]%
\>[28]{}(\Varid{insert}\;\Varid{v}\;\Varid{e}),\Varid{o}),{}\<[47]%
\>[47]{}(\Varid{xs},\Varid{ys}),{}\<[58]%
\>[58]{}\Conid{Leaf}\;\Varid{x}){}\<[E]%
\\
\>[B]{}\Varid{sortTree''}\;{}\<[13]%
\>[13]{}(\Conid{Leaf}\;\Varid{v})\;{}\<[24]%
\>[24]{}((\Varid{e},\Conid{Ordlist}\;{}\<[38]%
\>[38]{}\Varid{o})\; {}\<[45]%
\>[45]{}\mathord{\sim}((\Varid{xs},(\Varid{y},\Varid{ys})){}\<[E]%
\\
\>[B]{}\hsindent{3}{}\<[3]%
\>[3]{}\mid \Varid{odd}\;{}\<[11]%
\>[11]{}\Varid{v}{}\<[15]%
\>[15]{}\mathrel{=}((\Varid{e},\Conid{OrdList}\;{}\<[31]%
\>[31]{}(\Varid{insert}\;\Varid{v}\;\Varid{o})),{}\<[47]%
\>[47]{}(\Varid{xs},\Varid{ys}),{}\<[58]%
\>[58]{}\Conid{Leaf}\;\Varid{x}){}\<[E]%
\\[\blanklineskip]%
\>[B]{}\Varid{sortTree''}\;(\Conid{Bin}\;\Varid{l}\;\Varid{r})\;\Varid{rest}\;\Varid{xs}\mathrel{=}{}\<[E]%
\\
\>[B]{}\hsindent{4}{}\<[4]%
\>[4]{}\mathbf{let}\;{}\<[9]%
\>[9]{}(\Varid{vl},\Varid{xsl},\Varid{tl}){}\<[27]%
\>[27]{}\mathrel{=}\Varid{sortTree''}\;\Varid{l}\;\Varid{vr}\;\Varid{xs}{}\<[E]%
\\
\>[9]{}(\Varid{vr},\Varid{xsr},\Varid{tr}){}\<[27]%
\>[27]{}\mathrel{=}\Varid{sortTree''}\;\Varid{r}\;\Varid{rest}\;\Varid{xsl}{}\<[E]%
\\
\>[B]{}\hsindent{4}{}\<[4]%
\>[4]{}\mathbf{in}\;(\Varid{vl},\Varid{xsr},\Conid{Bin}\;\Varid{tl}\;\Varid{tr}){}\<[E]%
\ColumnHook
\end{hscode}\resethooks

Note that again we can provide all parameters at once,
and use guards.  We could have written this code in GHC style,
computing the types of the pair of   Cartesian products by first
inspecting not only the shape of the tree but also the values stored
in the tree. We think however that our approach, in which we see the
computed existential type as part of the result, is a more
natural one given that we are dealing with a language which has lazy
evaluation.

\section{Future work} 
With respect to the implementation of the \ensuremath{\Conid{ST}} monad, 
one may want to remark that the presented implementation is very
inefficient, since access to components of the state is done in
linear time. 
It is however possible to lazily convert the Cartesian product
into a tree-like structure, which gives us logarithmic lookup time. 
From this tree we may compute a list of indices in the tree, 
which can then be passed on from left-to-right through the computation. 
Whenever we add an element to the state, we take the first index from this list, 
since it is guaranteed to point
to the position in the tree-shaped state where the element currently
being added will end up.

There is still work to be done to mechanically verify 
the soundness of our type rules, in the sense that ``no well-formed
program can go wrong''. It is clear from our description that the type
rules will depend on the user providing sufficient type annotations,
since the standard HM inference system is not able to infer neither
the \ensuremath{\exists} nor the
\ensuremath{{\bar{\exists}}} quantifiers.

\section{Discussion}

We have completed our description. We think however that it is
desirable to spend some attention to why we managed to implement our
code, whereas it is not accepted by GHC.  There are several reasons.

In the first place the Utrecht Haskell Compiler allows to specify an existential type without the introduction of an extra intervening data type. 
This makes the code more concise, but is not essential. In GHC we could have defined our \ensuremath{\Conid{ST}} type by introducing an extra type \ensuremath{\Conid{ST'}}:
\begin{hscode}\SaveRestoreHook
\column{B}{@{}>{\hspre}l<{\hspost}@{}}%
\column{3}{@{}>{\hspre}c<{\hspost}@{}}%
\column{3E}{@{}l@{}}%
\column{6}{@{}>{\hspre}l<{\hspost}@{}}%
\column{10}{@{}>{\hspre}l<{\hspost}@{}}%
\column{18}{@{}>{\hspre}l<{\hspost}@{}}%
\column{20}{@{}>{\hspre}l<{\hspost}@{}}%
\column{22}{@{}>{\hspre}l<{\hspost}@{}}%
\column{32}{@{}>{\hspre}l<{\hspost}@{}}%
\column{39}{@{}>{\hspre}l<{\hspost}@{}}%
\column{47}{@{}>{\hspre}l<{\hspost}@{}}%
\column{49}{@{}>{\hspre}l<{\hspost}@{}}%
\column{E}{@{}>{\hspre}l<{\hspost}@{}}%
\>[B]{}\mathbf{data}\;\Conid{ST}\;{}\<[10]%
\>[10]{}\Varid{s}\;{}\<[18]%
\>[18]{}\Varid{a}{}\<[E]%
\\
\>[B]{}\hsindent{3}{}\<[3]%
\>[3]{}\mathrel{=}{}\<[3E]%
\>[20]{}\Conid{ST}\;\forall\;.\;\Varid{t}_{\Varid{rest}}{}\<[39]%
\>[39]{}\to \Varid{s}{}\<[47]%
\>[47]{}\to \Conid{ST'}\;\Varid{s}\;\Varid{t}_{\Varid{rest}}\;\Varid{a}{}\<[E]%
\\
\>[B]{}\mathbf{data}\;\Conid{ST'}\;\Varid{s}\;\Varid{t}_{\Varid{rest}}\;{}\<[20]%
\>[20]{}\Varid{a}{}\<[E]%
\\
\>[B]{}\hsindent{3}{}\<[3]%
\>[3]{}\mathrel{=}{}\<[3E]%
\>[6]{}\forall\;\Varid{t}_{\Varid{env}}\;.\;{}\<[22]%
\>[22]{}\Conid{ST'}\;{}\<[32]%
\>[32]{}(\Conid{T}\;\Varid{t}_{\Varid{env}}\;\Varid{s}){}\<[49]%
\>[49]{}\to (\Varid{a},\Varid{t}_{\Varid{env}},\Conid{T}\;\Varid{t}_{\Varid{rest}}\;\Varid{s}){}\<[E]%
\ColumnHook
\end{hscode}\resethooks

A more serious problem is that pattern matching for existential types in GHC is strict, 
and we thus cannot unpack \cite{Mitchell:1988:ATE:44501.45065} such a value in the right hand side of a \ensuremath{\mathbf{let}}. 
This restriction (probably) finds its roots in the fact that existential types naturally come with GADT's and 
that in the current GHC implementation non-strict pattern matching for GADT's may lead to unsafe code \cite{JFP:8368538}.  
There are two solutions for this. If the GADT does not 
introduce equality constraints, as is the case in our code, the restriction could be relieved. 
Another solution is to represent the equality constraints implicitly in the generated code. This corresponds to the approach taken in the pre-cursor of 
GADTs by Baars and Swierstra \cite{BaSw02}, where equality constraints are represented as coercions which 
are called when the proof that two types are equal is needed; failing to return such a proof leads to non-termination. 
The current GHC approach is thus more 
strict in requiring that it can be statically determined that a proof exists, and thus does not have to be checked for dynamically.
This restriction makes it impossible to pass the value back into the computation as demonstrated by a `rewrite' of \ensuremath{\Varid{runST}}  
where the passed parameter \ensuremath{\Varid{s}} has become unbound.
Pairing \ensuremath{\Varid{s}} with the result \ensuremath{\Varid{a}} in attempt to get hold of it, so we can feed it back, does not work either since the existential type in that case 
`escapes' \cite{JFP:8368538}.   
 \begin{hscode}\SaveRestoreHook
\column{B}{@{}>{\hspre}l<{\hspost}@{}}%
\column{18}{@{}>{\hspre}l<{\hspost}@{}}%
\column{20}{@{}>{\hspre}l<{\hspost}@{}}%
\column{34}{@{}>{\hspre}l<{\hspost}@{}}%
\column{E}{@{}>{\hspre}l<{\hspost}@{}}%
\>[B]{}\Varid{runST}\;(\Conid{ST}\;\Varid{st})\mathrel{=}{}\<[18]%
\>[18]{}\mathbf{case}\;\Varid{st}\;()\;\Varid{s}\;(\Conid{T}\;\Varid{id})\;\mathbf{of}{}\<[E]%
\\
\>[18]{}\hsindent{2}{}\<[20]%
\>[20]{}(\Varid{a},\Varid{s},\anonymous ,\anonymous )\to {}\<[34]%
\>[34]{}\Varid{a}{}\<[E]%
\ColumnHook
\end{hscode}\resethooks

Note that the latter problem could be solved by using a more fine-grained description of the use of existentials \cite{Montagu:2009:MAT:1480881.1480926}. 

But in all these cases the problem remains that conventional System-F does not allow for a \ensuremath{\mathbf{letrec}} construct at the type level:
we cannot use a type which we have gotten access to by unpacking it,
as a type parameter to a polymorphic function the call of which produced that very type. 

We finish by noting that the conventional translation of the use of existential types
into a polymorphic types as in:
\begin{hscode}\SaveRestoreHook
\column{B}{@{}>{\hspre}l<{\hspost}@{}}%
\column{E}{@{}>{\hspre}l<{\hspost}@{}}%
\>[B]{}\exists\;\Varid{t}\;\sigma\Rightarrow \forall\;\sigma'\;.\;(\forall\;\Varid{t}\;.\;\sigma\to \sigma')\to \sigma'{}\<[E]%
\ColumnHook
\end{hscode}\resethooks
does not apply to \ensuremath{{\bar{\exists}}}.
 
\section{Conclusion}
We have shown how to widen the use of existential types, such that besides
polymorphic functions we can too have polymorphic contexts. As the
main example of the usefulness of this approach we have given an
alternative implementation of the \ensuremath{\Conid{ST}} monad. A similar version was
developed in \cite{BSV09}. In that implementation a state was
constructed in a left-to-right fashion, with the constructed state
coming out at the right. This state had to be fed back in order to be
able to run the state. Unfortunately there the last added elements end up at
the beginning of the Cartesian product, so handed out references had
to be updated. This makes a monadic interface impossible, and instead an
arrow-based interface was given. We believe the implementation given
here is the one to be preferred, because of its more expressive interface.
In the original implementation of the TTTAS (Typed Transformations of
Typed Abstract Syntax)\cite{BSV09b, BSV09}
library we had to make provisions
for maintaining so-called meta information during the transformation process.
With the monadic interface this extra provision is no longer needed, 
and thus the library can be simplified considerably.

Although the \ensuremath{\Varid{idTree}} example may seem to be quite artificial we have
encountered the pattern used in there quite often. When programming in an
attribute-grammar based style (writing so-called \emph{circular}
functions) one gets very accustomed to constructing
values from trees, which are later fed back into the computation. Although
many of these applications could be rewritten into multi-visit
functions, this implies the explicit construction of intermediate
representations and makes resulting programs much more difficult to develop
and maintain. 

Another example where a constructed value is passed back
can be found in the implementation of a pretty printer which has a
bounded look-ahead \cite{CambridgeJournals:2837460}. In this algorithm
two processes walk over a tree-like structure which we want
to layout in a nice way. These processes communicate with each other
through streams, which we represented as lists. One process produces
a list of questions to be answered by the second process, which
communicates back these answers through another list.  The latter process produces
a list which is threaded backwards through the tree and which is,
when it emerges at the top, passed back into the tree and then passed on in a
left-to-right tree traversal. In the attribute grammar based
implementation \cite{PPTr2004} it is not enforced
that the first process adds exactly one element to the list of questions  
for each node of interest, nor that the second process produces exactly one answer for
each question, and that the first process consumes exactly one answer
for each question asked. Using the techniques described in this paper we
may enforce these requirements.

We finally want to remark that the code we have written is by no means special once one gets
used to the `data-flow' view of lazy functional programming.
Especially when trying to translate the results of an attribute grammar based development
--which are of a data flow view by nature-- into Haskell it is that one runs into the kind of problems we have addressed;
information flowing backwards and forwards is likely to occur in such developments 
and we argue that a type system should not make it impossible to express this in a type-safe way. 

\section{Acknowledgments}
We want to thank Andres L\"oh, Tom Schijvers and Andrea Vezzosi for discussing System-F related issues with us.


\bibliographystyle{splncs03}
\begin{flushleft}
\bibliography{thesis}
\end{flushleft}

\end{document}